\documentclass{aa}

\usepackage[tbtags,fleqn]{amsmath}
\usepackage{amsfonts}
\usepackage{amssymb}
\usepackage{pstricks, pst-plot}
\usepackage{graphicx}
\usepackage{natbib}
\usepackage{longtable}

\newcommand{\diff}{\mathrm d}

\newcommand{\mincir}{\raise
  -2.truept\hbox{\rlap{\hbox{$\sim$}}\raise5.truept \hbox{$<$}\ }}
\newcommand{\magcir}{\raise
  -2.truept\hbox{\rlap{\hbox{$\sim$}}\raise5.truept \hbox{$>$}\ }}


\begin{document}

\title{2MASS wide field extinction maps: I. The Pipe nebula}
\author{Marco Lombardi\inst{1,2}, Jo\~ao Alves\inst{1}, and Charles
  J.~Lada\inst{3}} \offprints{M.~Lombardi} \mail{mlombard@eso.org}
\institute{%
  European Southern Observatory, Karl-Schwarzschild-Stra\ss e 2,
  D-85748 Garching bei M\"unchen, Germany \and University of Milan,
  Department of Physics, via Celoria 16, I-20133 Milan, Italy \and
  Harvard-Smithsonian Center for Astrophysics, Mail Stop 42, 60 Garden
  Street, Cambridge, MA 02138}
\date{Received ***date***; Accepted ***date***}
\abstract{
}{
  We present a $8^\circ\times6^\circ$, high resolution extinction map
  of the Pipe nebula using 4.5 million stars from the Two Micron All
  Sky Survey (2MASS) point source catalog.
}{
  The use of \textsc{Nicer} (Lombardi \& Alves 2001), a robust and
  optimal technique to map the dust column density, allows us to
  detect a $A_V = 0.5 \mbox{ mag}$ extinction at a 3-$\sigma$ level
  with a $1 \mbox{ arcmin}$ resolution.
}{
  (\textit{i\/}) We find for the Pipe nebula a \textit{normal\/}
  reddening law, $E(J - H) = (1.85 \pm 0.15) E(H - K)$.
  (\textit{ii\/}) We measure the cloud distance using Hipparchos and
  Tycho parallaxes, and obtain $130^{+24}_{-58} \mbox{ pc}$.  This,
  together with the total estimated mass, $10^4 \mbox{ M}_\odot$,
  makes the Pipe the closest massive cloud complex to
  Earth. (\textit{iii\/}) We compare the \textsc{Nicer} extinction map
  to the NANTEN ${}^{12}$CO observations and derive with
  unprecedented accuracy the relationship between the near-infrared
  extinction and the $^{12}$CO column density and hence (indirectly)
  the $^{12}$CO X-factor, that we estimate to be $2.91 \times 10^{20}
  \mbox{ cm}^{-2} \mbox{ K}^{-1} \mbox{ km}^{-1} \mbox{ s}$ in the
  range $A_V \in [0.9, 5.4] \mbox{ mag}$. (\textit{iv\/}) We identify
  approximately $1\,500$ OH/IR stars located within the Galactic bulge
  in the direction of the Pipe field. This represents a
  significant increase of the known numbers of such stars in the
  Galaxy.
}{
  Our analysis confirms the power and simplicity of the color excess
  technique to study molecular clouds.  The comparison with the NANTEN
  ${}^{12}$CO data corroborates the insensitivity of CO observations to
  low column densities (up to approximately $2 \mbox{ mag}$ in $A_V$),
  and shows also an irreducible uncertainty in the dust-CO correlation
  of about $1 \mbox{ mag}$ of visual extinction. 
}
\maketitle

%

\defcitealias{2001A&A...377.1023L}{Paper~I}

\section{Introduction}
\label{sec:introduction}

Nearby Galactic molecular clouds complexes represent our best chance
to understand cloud formation and evolution and hence to learn how
stars come to be.  But progress on the study of these objects has been
slow.  Not only are molecular clouds the coldest objects known in the
Universe, their main mass component ($\mathrm{H}_2$) cannot be
detected directly.  Most of all we know today about their physical
properties has been derived through radio spectroscopy of
$\mathrm{H}_2$ surrogate molecules (CO, CS, $\mathrm{NH}_3$; e.g.,
\citealp{1999osps.conf....3B, 1999osps.conf...67M}) and more recently
through thermal emission of the dust grains inside these clouds
\citep{2000prpl.conf...59A, 2000ApJ...545..327J}.  The results
obtained using these techniques, especially the estimate of column
densities, are not always straightforward to interpret and are plagued
by several poorly constrained effects.  Moreover, although large scale
maps of entire molecular cloud complexes are now available
\citep{1998ApJS..115..241H, 2001ApJ...551..747S}, maps with sufficient
resolution and dynamic range to identify not only dense molecular
cores but also their extended environment are still not existent.
This wide view on molecular clouds is at present an obvious gap in our
understanding of the relation between the dense Interstellar Medium
(ISM) and star formation.

A straightforward and powerful technique to study molecular cloud
structures, pioneered by \citet{1994ApJ...429..694L} and known as the
Near-Infrared Color Excess method (\textsc{Nice},
\citealp{1998ApJ...506..292A}) makes use of the most reliable tracer
of $\mathrm{H}_2$ in these clouds: extinction by pervasive dust grains
in the gas.  The \textsc{Nice} method relies on near-infrared (NIR)
measurements of extinguished background starlight to derive accurate
line-of-sight estimates of dust column densities.  Depending on the
stellar richness and color properties of the background field this
technique can produce column density maps with spatial resolutions
down to $5 \mbox{ arcsec}$ \citep[e.g.][]{2001Natur.409..159A,
  2002osp..conf...37A} and with dynamic ranges more then an order of
magnitude larger than classical optical star count techniques.  This
novel view on molecular clouds is providing not only new information
on the physical structure of these object \citep{2001Natur.409..159A}
but also an insight into their chemical structure and the physical
properties of the dust grains, when combined to molecular line and
dust emission data \citep{1998A&A...329L..33K, 1999A&A...342..257K,
  1999ApJ...515..265A, 2001ApJ...557..209B, 2003ApJ...586..286L,
  2003A&A...399L..43B, 2003A&A...399.1073K}.

In part to make use of the wealth of NIR data provided by the Two
Micron All Sky Survey (2MASS; \citealp{1994ExA.....3...65K}) the
\textsc{Nice} method was further developed into an optimized
multi-band technique dubbed Near-Infrared Color Excess Revisited
(\textsc{Nicer}, \citealp{2001A&A...377.1023L}, hereafter
Paper~I). This generalization of \textsc{Nice} can, nevertheless, be
applied to any multi-band survey of molecular clouds.  Through use of
optimal combinations of colors, \textsc{Nicer} improves the noise
variance of a map by a factor of two when compared to \textsc{Nice}.
This unique property of \textsc{Nicer} makes it the ideal tool to
trace large scale distributions of low column density molecular cloud
material.  When applied to 2MASS data, \textsc{Nicer} dust extinction
maps trace not only the low column density regions ($A_V \simeq 0.5
\mbox{ mag}$) but have the dynamic range to identify dense molecular
cores by reaching cloud depths of $\sim 30 \mbox{ mag}$ of
extinction, corresponding to $8 \times 10^{22} \mbox{ protons} \mbox{
  cm}^{-2}$ \citep{L05}.

In this paper we present an extinction map of the Pipe nebula covering
$48 \mbox{ sq deg}$, computed by applying the \textsc{Nicer} technique
on 4.5 millions JHK photometric measurements of stars from the 2MASS
database.  The Pipe nebula is a poorly studied nearby complex of
molecular clouds.  The only systematic analysis of this region is the
one of \citet{1999PASJ...51..871O} who present a $\sim 27 \mbox{ sq
  deg}$ map in the $J = 1-0$ line of $^{12}$CO observed on a $4 \mbox{
  arcmin}$ grid, and smaller maps of selected regions in the $J = 1-0$
lines of $^{13}$CO and C$^{18}$O.  They estimate the total $^{12}$CO
mass to be $\sim 10^4 \mbox{ M}_\odot$ (for a cloud distance of $160
\mbox{ pc}$) and point out that star formation is only occurring on
Barnard~59, where the C$^{18}$O column density is the highest and where
they find a CO outflow.  Barnard~59 was also observed at $1300\
\mu\mbox{m}$ by \citet{1996A&A...314..258R} and a protostellar
candidate B59-MMS1 was found.  Two pre-main sequence stars associated
with Barnard~59 appear in the young binaries survey of
\citet{1993A&A...278...81R} and more recently in
\citet{2002AJ....124.1082K}.

This paper is organized as follows.  In
Sect.~\ref{sec:nicer-absorpt-map} we describe the technique used to
map the dust in the Pipe nebula and we present the main results
obtained.  In Sect.~\ref{sec:distance} we address the determination of
the cloud distance using Hipparcos data.  A statistical analysis and a
discussion of the bias introduced by foreground stars is presented in
Sect.~\ref{sec:statistical-analysis}.  We compare the CO observations
from \citet{1999PASJ...51..871O} with our outcome in
Sect.~\ref{sec:comparison-with-co}.  Section~\ref{sec:mass-estimate}
is devoted to the mass estimate of the cloud complex.  Finally, we
summarize the conclusions of this paper in
Sect.~\ref{sec:conclusions}.

In the following we will normally express column densities in terms of
the 2MASS $K_\mathrm{s}$ band extinction $A_K$.  When converting this
quantity into the widely used visual extinction $A_V$, we will use the
\citet{1985ApJ...288..618R} reddening law converted into the 2MASS
photometric system \citep[see][]{2001AJ....121.2851C}, $A_K / A_V =
0.112$.

\section{\textsc{Nicer} extinction map}
\label{sec:nicer-absorpt-map}

\begin{figure}[!t]
  \centering
  \includegraphics[width=\hsize]{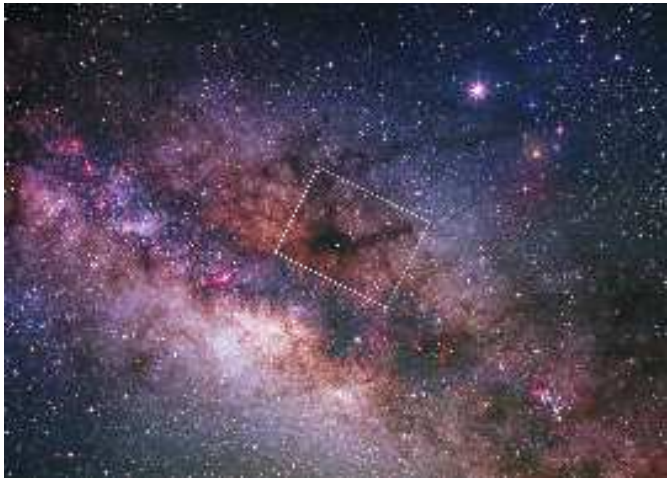}
  \caption{Color optical image of the region around the Pipe nebula.
    The bright ``star'' on the top-right is Jupiter; below it are well
    visible the $\rho$-Ophiuci dark cloud and the Lupus complex.  The
    white box encloses the field studied in this paper.  The Pipe
    nebula obscures part of the Galactic bulge, and thus occupies an
    exceptional location for infrared studies (image courtesy Jerry
    Lodriguss, \texttt{http://www.astropix.com/}). }
  \label{fig:1}
\end{figure}

The data analysis was carried out using the \textsc{Nicer} method
described in Paper~I (to which we refer for more detailed
information).  Infrared $J$ ($1.25\ \mathrm{\mu m}$), $H$ ($1.65 \
\mathrm{\mu m}$), and $K_\mathrm{s}$ band ($2.17 \ \mathrm{\mu m}$)
magnitudes of stars in the Pipe region were obtained from the Two
Micron All Sky Survey\footnote{See
  \texttt{http://www.ipac.caltech.edu/2mass/}.}
\citep[2MASS;][]{1994ExA.....3...65K}.  We selected a large region
around the Pipe nebula, characterized by galactic coordinates
\begin{align}
  \label{eq:1}
  -4^\circ <{} & l < +4^\circ \; , & +2^\circ <{} & b < +8^\circ \; .
\end{align}
As shown by these coordinate ranges, the Pipe nebula occupies an
exceptionally good location for infrared studies.  Indeed, this cloud
complex is just above the galactic plane and is in front of the
galactic bulge (see Fig.~\ref{fig:1}).  The cloud, we anticipate, is
approximately at $130 \mbox{ pc}$ (see Sect.~\ref{sec:distance}), and
thus almost all the stars observed in direction of the cloud are
background objects.  As a result, we could carry out the study of the
infrared extinction of the cloud with negligible contamination by
foreground objects (except in the most dense regions).

The high density of stars in the Pipe region has also some drawbacks.
Indeed, because of confusion, the nominal 2MASS photometric
completeness limits drop close to the galactic center.  In the case of
\textsc{Nicer} (in contrast with the star-counting method), this does
not affect the measurements of column densities because no assumptions
are made about the local star density.  High density regions also pose
severe challenges for data reduction.  For example, the pipeline used
for the Second Incremental Release of 2MASS produced slightly
inaccurate zero-points.  This problem showed on the \textsc{Nicer}
maps of the Pipe nebula as abrupt artificial changes in the measured
column density at the boundaries of the 2MASS observation tiles.
Fortunately, the pipeline has been greatly improved and this problem
does not appear in the final 2MASS All Sky Release.

\begin{figure*}[!tbp]
  \begin{center}
    \includegraphics[bb=17 52 594 450,width=\hsize]{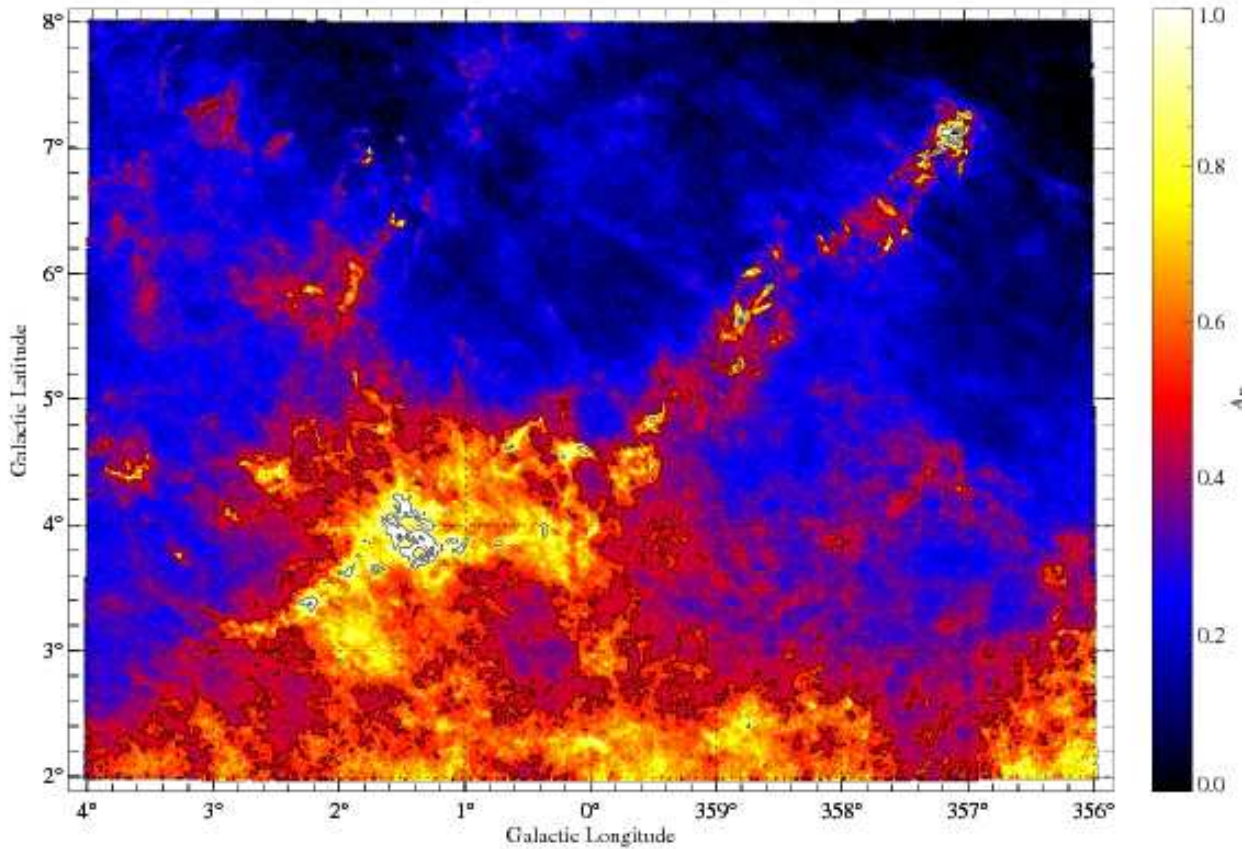}
    \caption{The \textsc{Nicer} extinction map of the Pipe nebula.
      The resolution is $\mathrm{FWHM} = 1 \mbox{ arcmin}$; the
      contours are at $A_K = \{0.5, 1, 1.5\}$ mags.}
    \label{fig:7}
  \end{center}
\end{figure*}

After selecting point sources from the 2MASS catalog inside the
boundaries \eqref{eq:1}, we generated a preliminary extinction map.
As described in Paper~I, this map was mainly used as a first check of
the data, to select a control region on the field, and to obtain there
the photometric parameters to be used in the final map (see
Fig.~\ref{fig:7}).  We note that the search of a control field close
to the Pipe nebula has been non-trivial, because of the complex cloud
structure and high column densities observed at low galactic
latitudes.  We identified in the Eastern part of our field (top-right
in Fig.~\ref{fig:7}) a small region that is apparently affected by
only a negligible extinction (see below); note that the newly
determined \citet{2005ApJ...619..931I} 2MASS reddening law was used
at this stage.

\begin{figure}[!t]
  \centering
  \includegraphics[width=\hsize]{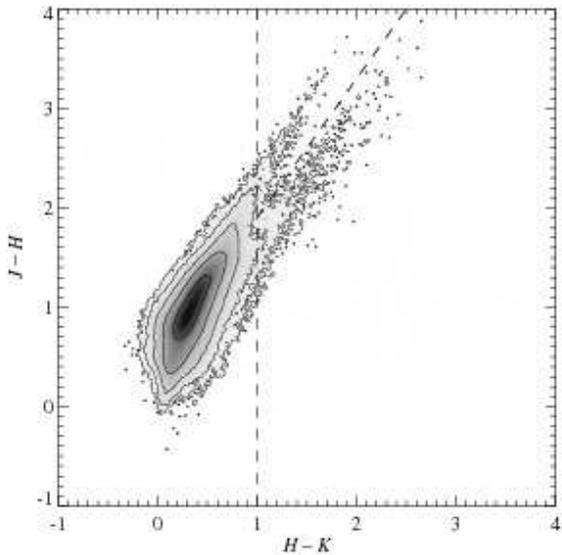}
  \caption{Color-color diagram of the stars in the Pipe nebula field,
    as a density plot.  The contours are logarithmically spaced, i.e.\
    each contours represents a density ten times larger than the
    enclosing contour; the outer contour detects single stars and
    clearly shows a bifurcation at large color-excesses.}
  \label{fig:2}
\end{figure}

Using the information provided by the control field, we generated
a second map, which is thus ``calibrated'' (i.e., provides already,
for each position in our field, a reliable estimate of the column
density).  We then considered the color-color diagram for the stars in
the catalog.  The result, shown in Fig.~\ref{fig:2}, shows a
surprising bifurcation for $H - K > 1 \mbox{ mag}$, which in principle
might represents a problem in color-excess studies.

\begin{figure}[!t]
  \centering
  \includegraphics[width=\hsize]{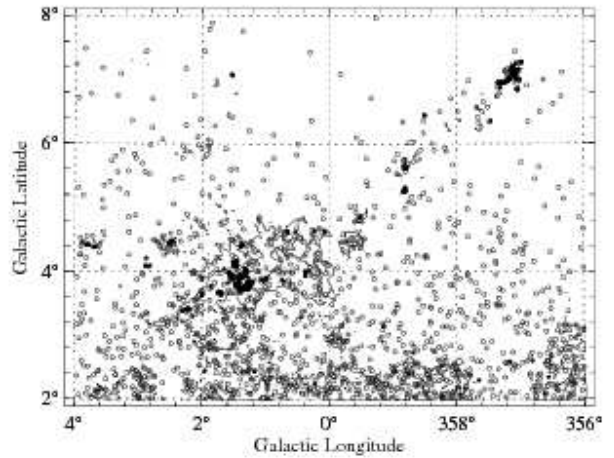}
  \caption{Spatial distribution of the subsample of sources as defined
    by Eqs.~\eqref{eq:2} and \eqref{eq:3}.  Subsample $A$ is
    shown as filled circles, while subsample $B$ is shown as open
    circles (see also Fig.~\ref{fig:2}).  Subsample $A$ appears to be
    strongly clustered in high-column density regions of the cloud,
    and are thus interpreted as genuine reddened stars; subsample $B$
    seems not to be associated with the cloud, and are instead
    preferentially located at low galactic latitudes.  The contour
    line represents the $A_K = 0.6$ contour of the Pipe nebula.}
  \label{fig:3}
\end{figure}

In order to further investigate the origin of this bifurcation, we
considered the region $H - K > 1 \mbox{ mag}$ in the color-color
diagram, and divided it into two subsamples $A$ (upper branch) and $B$
(lower branch) according to the expressions
\begin{align}
  \label{eq:2}
  A \equiv {} & \{ 1.4 (H - K) + 0.5 \mbox{ mag} < (J - H) \} \; , \\
  \label{eq:3}
  B \equiv {} & \{ 1.4 (H - K) + 0.5 \mbox{ mag} > (J - H) \} \; .
\end{align}
We then analysed the spatial distribution of the stars contained in
the two regions.  The results, shown in Fig.~\ref{fig:3}, give a first
strong indication that, as expected from their location in the
color-color diagram, subsample $A$ is associated with the densest
regions of the molecular cloud and thus it has to be interpreted as
normal stars reddened by the dust, while subsample $B$ does not appear
to be associated with the cloud.  Interestingly, subsample $B$ is
uniformly distributed in galactic longitude, but has a strong
preference for low galactic latitude regions.

\begin{figure}[!t]
  \centering
  \includegraphics[bb=141 298 456 531, width=\hsize]{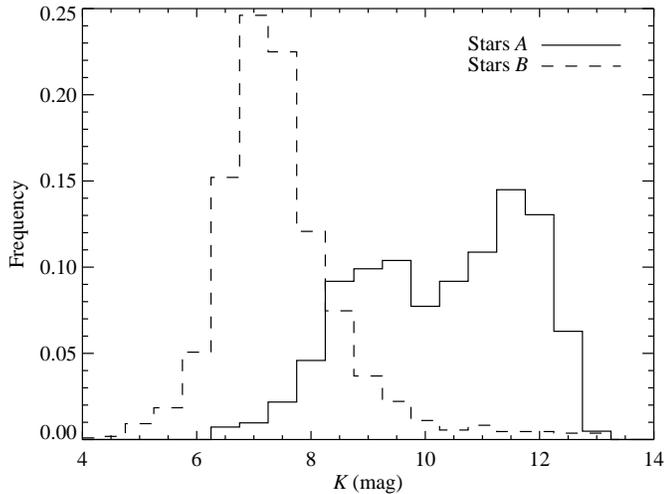}
  \caption{The histogram of the $K$ band magnitude for the two star
    subsets $A$ and $B$ of Eqs.~\eqref{eq:2} and \eqref{eq:3}.}
  \label{fig:4}
\end{figure}

Further hints on the nature of the two stellar populations in
subsamples $A$ and $B$ are given by the histogram of their $K$ band
magnitudes, shown in Fig.~\ref{fig:4}.  Interestingly, while subsample
$A$ stars show, as expected, a broad distribution, with number counts
increasing at relatively large magnitudes, the $B$ stars show a well
defined distribution, with a strong peak at $K \simeq 7 \mbox{ mag}$.

\begin{figure}[!t]
  \centering
  \includegraphics[width=\hsize]{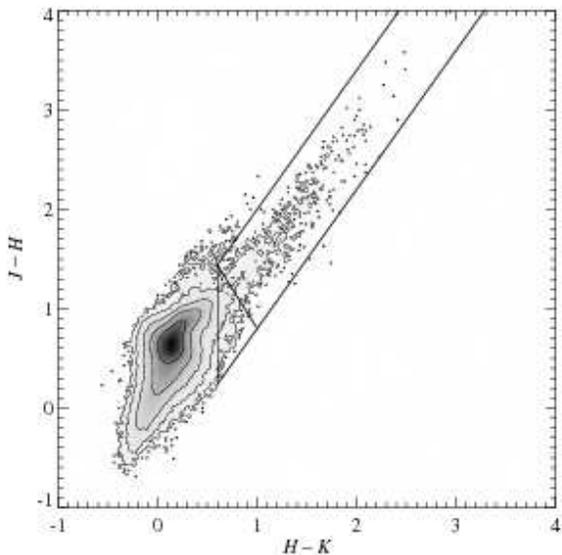}
  \caption{The extinction-corrected color-color diagram for the Pipe
    nebula.  The stripes show both subsamples $B_1$ and $B_2 \subset
    B_1$ [cf.\ Eqs.~\eqref{eq:4} and \eqref{eq:5}].}
  \label{fig:5}
\end{figure}

The lack of correlation between the dust reddening and the stars of
subsample $B$ is also well expressed by extinction-corrected
color-color diagram, shown in Fig.~\ref{fig:5}.  This plot has been
constructed by estimating, for each star of Fig.~\ref{fig:2}, its
\textit{intrinsic\/} color, obtained by correcting the observed one by
the measured \textsc{Nicer} extinction at the location of the star
(cf.\ Fig.~\ref{fig:7}).  This analysis, based on the simplifying
assumptions that all stars are background to the molecular cloud and
that the extinction is basically constant on the resolution of our
map, is however able to capture the essential characteristics of the
distribution of \textit{intrinsic\/} star colors.  By comparing
Fig.~\ref{fig:5} with Fig.~\ref{fig:2}, in particular, we see that the
upper branch, subsample $A$, essentially disappears in the
extinction-corrected plot, while the lower branch, subsample $B$, is
largely left unchanged.  Because of the much shrunken distribution of
stars, Fig.~\ref{fig:5} is particularly useful to better identify
stars belonging to the lower branch.  We defined thus in this plot
two sets
\begin{align}
  \label{eq:4}
  B_1 \equiv {} & \bigl\{ \bigl\lvert 1.4 (H - K)_\mathrm{intr} - (J -
  H)_\mathrm{intr} \bigr\rvert < 0.6 \notag\\
  & \phantom{\bigl\{} \mbox{ and } (H - K)_\mathrm{intr} > 0.6 \} \; ,
  \\
  \label{eq:5}
  B_2 \equiv {} & \bigl\{ \bigl\lvert 1.4 (H - K)_\mathrm{intr} - (J -
  H)_\mathrm{intr} \bigr\rvert < 0.6 \notag\\
  & \phantom{\bigl\{} \mbox{ and } (J - H)_\mathrm{intr} > -1.6 (H -
  K)_\mathrm{intr} + 2.4 \} \; .
\end{align}
Note that $B_2 \subset B_1$.  These two sets, marked in
Fig.~\ref{fig:5}, correspond to the areas in the color-color plot
where a contamination by subsample $B$ stars is possible ($B_1$), or
highly likely ($B_2$).

From the elements considered so far, we can carry out the following
conclusions with respect to the nature of the $B$ subsample that from
this point on will be called the ``lower branch'':
\begin{itemize}
\item These stars are unrelated to the molecular cloud reddening and
  appear to be located close to the galactic plane.  A deeper
  analysis, carried out on a much larger field (Lombardi et al., in
  prep.), actually shows that the stars are preferentially located on
  the galactic bulge.
\item The stars occupy a well defined region in the
  (extinction-corrected) color-color diagram.  This indicates that the
  ``lower branch'' represents a well-defined population of stars.
\item The $K$ band distribution of ``lower branch'' stars is strongly
  peaked at $K \simeq 7 \mbox{ mag}$.  This not only is a further
  indication that we are looking at a homogeneous population, but also
  suggests that these stars are at essentially the same
  \textit{distance\/}. The $K$-band standard deviation is
  approximately $0.8 \mbox{ mag}$ (if the tail at $K > 10 \mbox{ mag}$
  of Fig.~\ref{fig:4} is neglected), and since this must account for
  the scatter in the absolute magnitude, the scatter in the distance,
  and the photometric errors, it is likely that these stars have a
  relative scatter in their distance of the order of $25\%$ or less.
  Again, this is a further hint that the ``lower branch'' stars might
  be a bulge population.
\item The distribution of ``lower branch'' stars in the color-color
  diagram (Fig.~\ref{fig:2}) forms an elongated region parallel to
  reddening vector.  This shows that at least part of the large
  scatter in their intrinsic colors might be due to reddening, local
  or along the line of sight.
\end{itemize}

The items above suggest that the ``lower branch'' might be populated
by evolved intermediate mass stars (about 1 to 7 M$_\odot$), and
likely on the Asymptotic Giant Branch (AGB), at about the distance to
the Galactic center.  We find a similarity between the ``lower
branch'' stars and the sample of Galactic OH/IR stars in
\citet{2005A&A...431..779J}.  These authors conclude that the
luminosity and color distribution observed could be explained as the
result of a combined emission of a cool star ($T \sim 2500 \mbox{ K}$)
and a much cooler dust shell ($T < 800 \mbox{ K}$). The main effect of
the shell in the near-infrared colors would be to increase the
circumstellar reddening, which, together with differential
interstellar extinction effects from source to source, could explain
the width of the distribution observed.  Although we will not explore
further in this paper the nature of these stars we note that they are
particularly interesting objects, and given the large number of
identifications in our fields (we have approximately $45\,000$ $B_1$
stars and $1\,500$ $B_2$ ones) a dedicated study is warranted.

The ``lower branch'' stars seem to be unrelated to the molecular
cloud, but their colors would be interpreted by the \textsc{Nicer}
algorithm as a sign of extinction.  This, clearly, could in principle
bias our results toward a higher extinction, especially at low
galactic latitude regions.  However, we argue that the bias introduced
by ``lower branch'' stars is negligible.  Indeed, the density of these
objects is, in the worse case (set $B_1$) only $1\%$ of the average
density of stars, so that typically we have at most one ``lower
branch'' star contaminating each pixel, and since we have $\sim 30$
stars per pixels, the effects of ``lower branch'' stars is negligible
everywhere.

\begin{figure}[!t]
  \centering
  \includegraphics[width=\hsize]{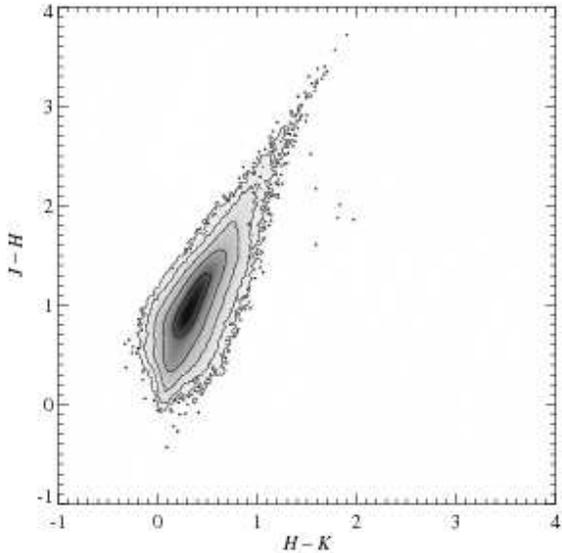}
  \caption{The color-color diagram for the selected stars in the
    field, after the removal of the set $B_2$ of Eq.~\eqref{eq:5}.  A
    comparison with Fig.~\ref{fig:2} shows that we were able to
    virtually remove all significant contamination from spurious
    reddening.}
  \label{fig:6}
\end{figure}

Nevertheless, and in order to avoid any source of bias, although
small, we excluded from the 2MASS catalogs all stars located in the
$B_2$ color-space region, and performed the whole analysis described
in this paper using this reduced subset of stars.  We stress that if
we had performed a a cut in the \textit{observed\/} colors, we would
be have introduced a new bias in the deduced column density; instead,
the use of a select in the \textit{intrinsic\/} colors does not bias
the final results.  As an example, Fig.~\ref{fig:6} shows the
color-color diagram for the new set of stars: note that the ``lower
branch'' disappears completely in this plot, a further confirmation
that our selection is effective in removing this population of stars.

\begin{figure*}[!tbp]
  \begin{center}
    \includegraphics[bb=17 52 594 450,width=\hsize]{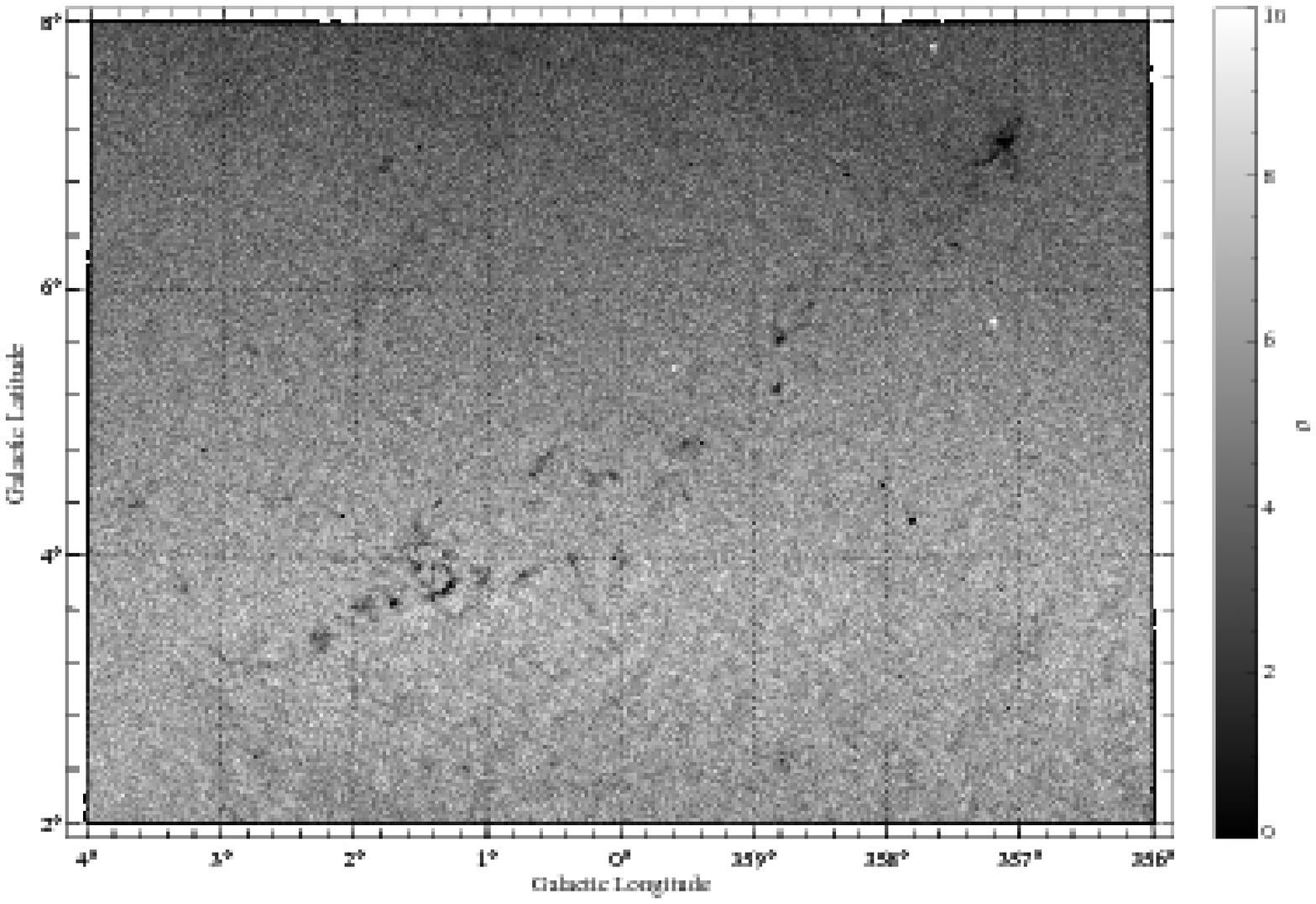}
    \caption{The star density map for the extinction map of
      Fig.~\ref{fig:7}, i.e.\ the number of stars inside a (Gaussian)
      $1 \mbox{ arcmin}$ beam.  The dark structures correspond mostly
      to the position of dense cores in the Pipe nebula where the star
      density in minimal due to high extinction. The black ``holes''
      outside the Pipe body correspond to masked bright stars while
      the white spots correspond to globular clusters.}
    \label{fig:8}
  \end{center}
\end{figure*}

We then run again the whole \textsc{Nicer} pipeline on the refined
catalog.  After (re)evaluating the statistical properties of stars in the
control field, we constructed the final map, shown in
Fig.~\ref{fig:7}, in a grid of approximately $1\,000 \times 750$ points,
with scale $30 \mbox{ arcsec}$ per pixels, and with Gaussian smoothing
characterized by $\mbox{FWHM} = 1 \mbox{ arcmin}$; moreover, we used
an iterative $\sigma$-clipping at $3$-$\sigma$ error.  The final,
\textit{effective\/} density of stars is $\sim 8$ stars per pixel
(this value changes significantly on the field with the galactic
latitude, see Fig.~\ref{fig:8}); this guarantees an average
($1$-$\sigma$) error on $A_K$ of only $0.019$ magnitudes; the largest
extinction is measured close to Barnard~59, where $A_K \simeq 2.68
\mbox{ mag}$ (corresponding to approximately $A_V \simeq 24 \mbox{
  mag}$).  As clearly shown by Fig.~\ref{fig:7}, the combination of
the use of the 2MASS archive with the optimized \textsc{Nicer}
technique allows us to reveal an unprecedented number of details.  A
quantitative analysis of this extinction map is delayed until
Sect.~\ref{sec:statistical-analysis}.

\begin{figure*}[!tbp]
  \begin{center}
    \includegraphics[bb=17 52 594 450,width=\hsize]{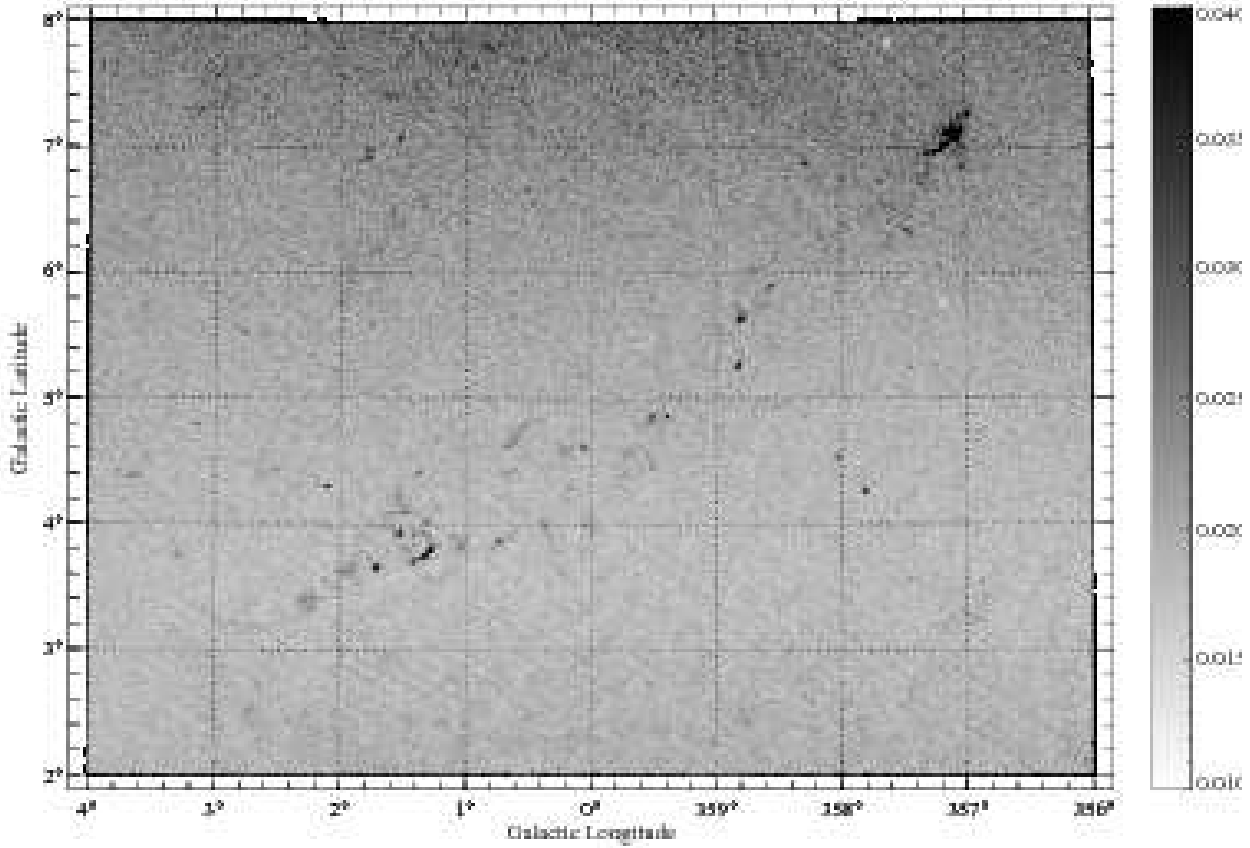}
    \caption{The map shows the statistical error $\sigma_{A_K}$ on the
      measured column density.  Note that correlation on the errors on
      a scale of $\mathrm{FWHM} = 1 \mbox{ arcmin}$ is expected.  This
      map shows that for almost the whole field $A_K = 0.05$ represents
      a 3-$\sigma$ detection.}
    \label{fig:9}
  \end{center}
\end{figure*}

\begin{figure}[!tbp]
  \begin{center}
    \includegraphics[bb=112 307 443 505, width=\hsize]{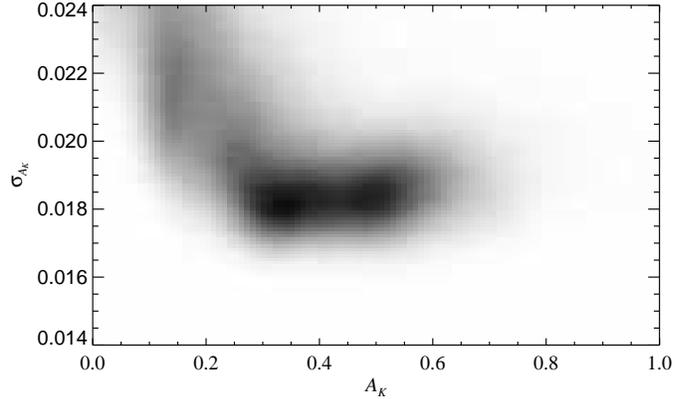}
    \caption{The expected error $\sigma_{A_K}$ versus the measured
      extinction $A_K$ for the various pixels in the extinction map of
      Fig.~\ref{fig:7}, shown as a density plot.}
    \label{fig:10}
  \end{center}
\end{figure}

Figure~\ref{fig:9} shows the expected error on $A_K$ for each pixel of
the extinction map.  This figure deserves a few comments.  First, we
note that the most significant variations in the expected error are
due to bright stars, which produce the characteristic cross-shaped
patterns (cf.\ Fig.~\ref{fig:8}).  We observe that, with the help of
Fig.~\ref{fig:9}, some apparent features in the extinction map are
actually recognized to be due to the larger noise expected close to
bright stars (these ``features,'' hence, are just increased
statistical deviations due to the increased local error).  This
clearly shows that a detailed analysis of the extinction map of
Fig.~\ref{fig:7} is better carried out using also the density map of
Fig.~\ref{fig:8} and the error map of Fig.~\ref{fig:9}.
Figure~\ref{fig:9} shows also an increase in noise in the densest
regions of the cloud, due to the reduced density of background stars.
Finally, we note the increase of the error with the galactic latitude,
also due to a change in the density of background stars.

The accuracy of the column density measurements obtained in our field
is shown by Fig.~\ref{fig:10}, which plots the expected error on $A_K$
as a function of $A_K$ for all pixels in our field.  The exquisite
data used allowed us to keep the average error well below $0.02 \mbox{
  mag}$ in $A_K$, and still to have a $1 \mbox{ arcmin}$ resolution in
our maps.  The dynamical range of the \textsc{Nicer} extinction map
can be better appreciated by noting that the lowest contour in
Fig.~\ref{fig:7} represents a $\sim 10 \sigma$ detection, and that
clumps such as Barnard~59 have a significance as large as $\sim 200
\sigma$.  Note that the increase on the error observed for $A_K < 0.2
\mbox{ mag}$ in Fig.~\ref{fig:10} is due to the fact that regions with
low extinction (including the control field) are located at high
galactic latitudes, and hence have a smaller density of background
stars.

\section{Distance}
\label{sec:distance}

An accurate determination of the distance of molecular clouds is of
vital importance to obtain a reliable estimate of the mass and of
other physical properties.  Unfortunately, distance measurements of
molecular clouds are frequently plagued by very large uncertainties.
A simple method used often is based on the association between the
cloud and other astronomical objects whose distance is well known.
\citet{1999PASJ...51..871O} associate the Pipe nebula with the
Ophiuchus complex on the base of projected proximity and radial
velocity and use the distance of the latter, $(160 \pm 20) \mbox{
  pc}$, as the distance to the Pipe \citep{1981A&A....99..346C}.

An alternative approach is based on the number counts of foreground
stars.  The method, used for example in \citet{1998ApJ...506..292A},
exploits the large reddening produced by some clouds, which makes the
identification of foreground stars relatively easy; then, galactic
models \citep[e.g.][]{1980ApJS...44...73B,1992ApJS...83..111W} are
used to infer the expected number of stars (for each possible cloud
distance) inside the cone created by the cloud.  Finally, the number
of foreground stars observed is compared to the prediction and the
distance of the cloud is inferred.  Although this method is often the
best one can use, it is unable to give accurate distances for several
reasons: (i) it relies on galactic models, which might be inaccurate
(especially at the small angular scales often used for molecular
clouds); (ii) it is plagued by Poisson noise (because the number of
stars inside the volume of the cone is a random variable) and (iii)
stars do cluster (and thus the error is actually larger than the one
expected from a Poisson statistics; imagine, in the extreme, the case
of an unknown open cluster in front of the cloud).

A more robust determination of the distance of the Pipe molecular
complex can be obtained using the Hipparcos and Tycho catalogs
\citep{1997A&A...323L..49P}.  The method, similarly to the star number
counts described in the previous section, is based on the
identification of foreground and background stars (observed on the
line of sight of the cloud) for which a parallax estimate is
available.  An upper limit for the distance of the cloud is thus given
by the distance of the closest background star, i.e.\ the closest star
showing a significant extinction in its colors.  This novel approach
to the distance of molecular cloud complexes has already been
successfully applied to several clouds by \citet{1998A&A...338..897K}.
Here we revisit the method and use it to obtain a distance estimate
for the Pipe nebula.

\begin{figure}[!tbp]
  \begin{center}
    \includegraphics[bb=131 290 454 540, width=\hsize]{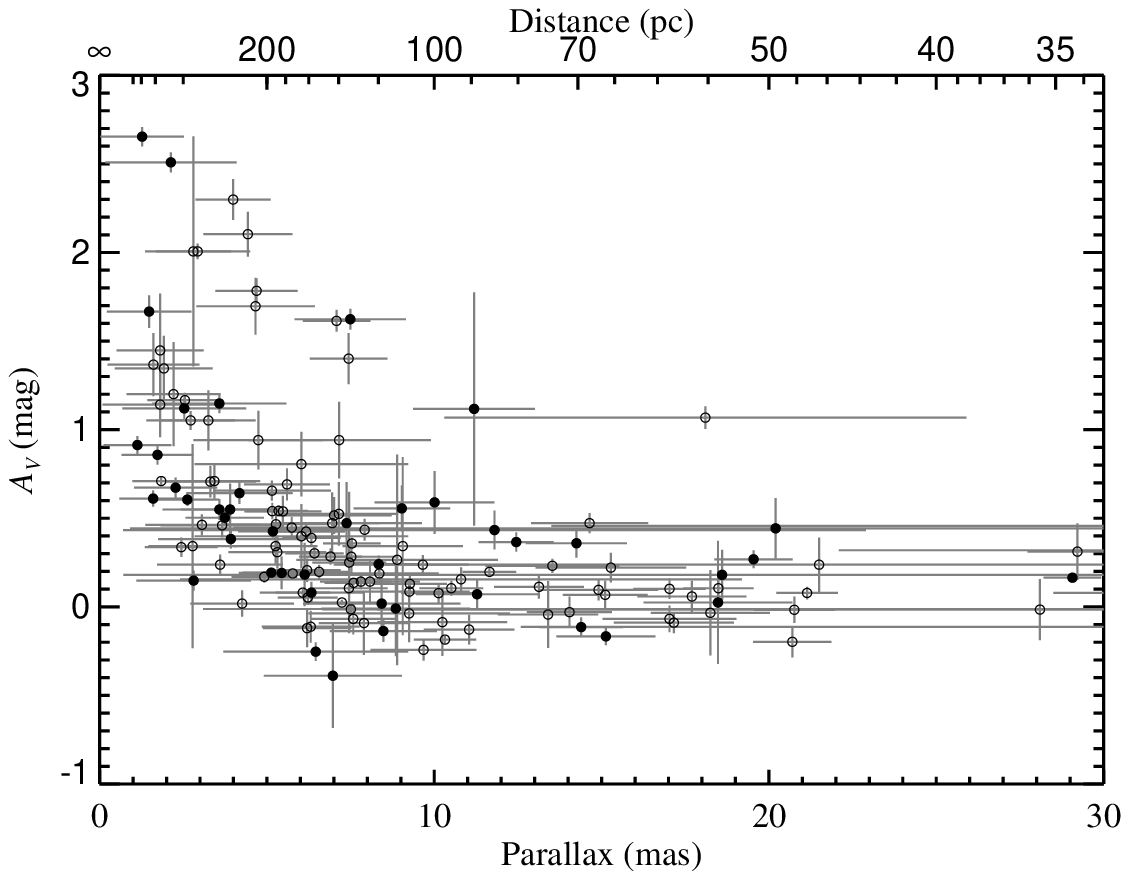}
    \caption{The reddening of Hipparcos and Tycho stars in the field.
      For this plot we selected only the stars characterized by a
      measured parallax $\hat\pi$ smaller than its error
      $\hat\sigma_\pi$, and have an estimated error on the column
      density $\hat\sigma_{A_V} < 1 \mbox{ mag}$.  Among the 151 stars
      that matches these constraints, we marked with a filled dots the
      44 ones that are located at $b > 3^\circ$.}
    \label{fig:11}
  \end{center}
\end{figure}

We selected Hipparcos and Tycho-1 (I/239) stars observed in the area
defined in Eq.~\eqref{eq:1}, and matched these stars with the
``All-sky Compiled Catalog of 2.5 million stars'' (ASCC-2.5, I/280A;
\citealp{2001KFNT...17..409K}) and with the ``Tycho-2 Spectra Type
Catalog'' (Tycho-2spec, III/231; \citealp{2003AJ....125..359W}).  The
choice of these two latter datasets (which are cross-references and
merges of many stellar catalogs) was dictated by the need to obtain
for each star with measured parallax an estimate of its spectral type.
In particular, the use of ASCC-2.5 and of Tycho-2spec effectively
allowed us to cover a large sample of spectroscopic catalogs: the
Hipparcos catalog (I/239), the Carlsberg Meridian Catalogs (CMC11;
I/256), the Position and Proper Motions (PPM; I/146, I/193, I/208),
the Michigan Catalogs (III/31, III/51, III/80, III/133, III/214), the
Catalog of Stellar Spectra Classified in the Morgan-Keenan System
(III/18), the MK Classification Extension (III/78), and the FK5
catalog parts I and II (I/149 and I/175).

We considered all stars in our field with measured parallax larger
than the parallax error, and with spectral type B, A, F, G, K, or M.
By comparing the expected $B - V$ color (taken from \citealp{LB},
p.~15) with the observed one, we obtained an estimated of the color
excess $E(B - V)$; we finally converted this into an extinction in the
$V$ band $A_V$ by using a normal reddening law, $A_V = 3.09 E(B-V)$
\citep{1985ApJ...288..618R}.  Note that, whenever possible, we used 2D
spectral types; for 1D spectral types we assumed a luminosity class
IV.

A plot of the star column density versus the Hipparcos and Tycho2
parallaxes is shown in Fig.~\ref{fig:11}.  Because of the relative
large scatters in the parallax and column density measurements,
Fig.~\ref{fig:11} is not straightforward to interpret.  However, we
note the following points
\begin{enumerate}
\item At large parallaxes, for $\pi > 12 \mbox{ mas}$, the measured
  extinction is low and consistent, within the errors, with a constant
  value.  The average extinction in this range, $\langle A_V \rangle =
  0.12 \mbox{ mag}$, either suggests the presence of a thin ``veil''
  very close to us or of a small systematic error in the estimate of
  $A_V$.  The presence of absorbing material towards the location of
  the Pipe nebula at distances below $50 \mbox{ pc}$ is also suggested
  by the recent work of \citet{2003A&A...411..447L}, where they
  construct 3D absorption maps of the local distribution of neutral
  gas through measurements of equivalent widths of the interstellar
  NaI D-line doublet towards nearby stars.  In any case, from
  Fig.~\ref{fig:11} we can deduce a robust lower limit on the cloud
  distance, $d > 80 \mbox{ pc}$.  Note that, interestingly, we also
  detect a near single star (HD~158233) at $(55.0 \pm 1.67) \mbox{
    mas}$ with significant reddening, $A_V = (0.38 \pm 0.06) \mbox{
    mag}$: thus, if the reddening has to be attributed to a thin
  molecular cloud, its distance would be smaller than $20 \mbox{ pc}$.
\item As we go to smaller parallaxes we start seeing stars with signs
  of a possible reddening.  A first candidate is a star at $90 \mbox{
    pc}$ with $A_V = (1.11 \pm 0.66) \mbox{ mag}$ (well visible in
  Fig.~\ref{fig:11}); however, the large error on $A_V$ does not allow
  us to assess securely that the star has been significantly reddened
  by the cloud.
\item At approximately $\pi = 7.3 \mbox{ mas}$ ($\sim 137 \mbox{ pc}$)
  we observe four stars with relatively large extinction, $A_V > 0.9
  \mbox{ mag}$, and with relative small errors (the ratio $A_V /
  \mathrm{Err}(A_V)$ ranges from $4.3$ to $27.5$).  Hence, these stars
  give an upper limit to the cloud distance, $d < 140 \mbox{ pc}$.
\end{enumerate}
In summary, the simple argument discussed above suggests that the Pipe
nebula is located between $80$ and $140 \mbox{ pc}$.  

\begin{figure}[!tbp]
  \begin{center}
    \includegraphics[bb=123 290 455 540, width=\hsize]{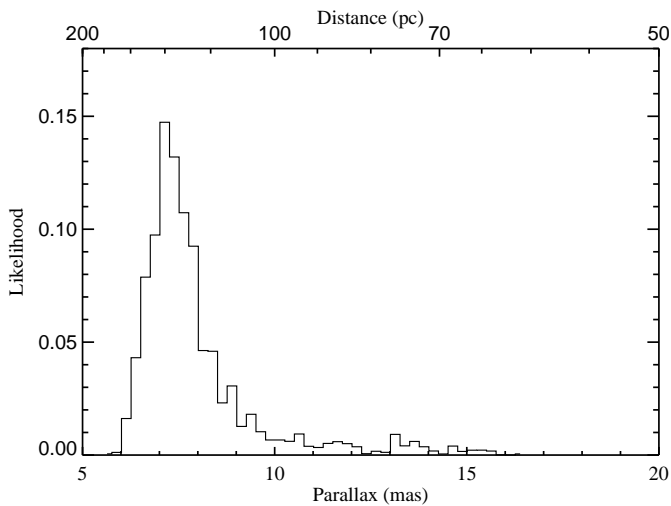}
    \caption{The likelihood function of Eq.~\eqref{eq:7} as a function
      of the cloud parallax $\pi_\mathrm{Pipe}$ marginalized over all
      the other parameters.  The function shows a marked peak at $\pi
      \simeq 7.13 \mbox{ mas}$.  The mean of $\pi_\mathrm{Pipe}$ is
      $8.07 \mbox{ mas}$, while the median is $7.61 \mbox{ mas}$,
      corresponding to a distance of approximately $130 \mbox{ pc}$.}
    \label{fig:12}
  \end{center}
\end{figure}

A more rigorous estimate of the distance can be obtained by using a
statistical model for the parallax-extinction relation.  The physical
picture here suggests that we can take the star column density as a
random variable whose distribution only depends on the intrinsic
measurement errors and on the parallax of the star.  The simplest
sensible approach is to consider a bimodal distribution for each star
column density, where the extinction of stars in front of the nebula
is consistent (within the measurement errors) with zero, and the
extinction of stars at distances larger than the cloud can be either
still zero (if the star is not observed through the cloud) or finite
and with a large scatter (if the star is behind the cloud).  In this
case, the conditional probability of observing a star with visual
extinction $A_V$ given the fact that the star is located at a parallax
$\pi$ is
\begin{equation}
  \label{eq:6}
  p(A_V | \pi) =
  \begin{cases}
    \mathrm{Gau}(A_V | A_V^\mathrm{fg}, 
    \sigma_{A_V}^{\mathrm{fg}2} + \sigma_{A_V}^2) & 
    \text{if $\pi > \pi_\mathrm{Pipe} \; ,$} \\[0.5em]
    (1 - f) \mathrm{Gau}(A_V | A_V^\mathrm{fg},
    \sigma_{A_V}^{\mathrm{fg}2} + \sigma_{A_V}^2) & \\
    {} + f \mathrm{Gau}(A_V | A_V^\mathrm{bg},
    \sigma_{A_V}^{\mathrm{bg}2} + \sigma_{A_V}^2) & 
    \text{if $\pi \le \pi_\mathrm{Pipe} \; ,$}
  \end{cases}
\end{equation}
where we denoted with $\mathrm{Gau}(x | \bar{x}, \sigma_x)$ the value
at $x$ of a normal probability density with mean $\bar{x}$ and
variance $\sigma^2_x$ [cf.\ Eq.~\eqref{eq:12}] .  In other words, we
use a normal distribution with low average $A_V^\mathrm{fg}$ for
foreground stars, and the \textit{mixture\/} of the same distribution
with another normal (with large average $A_V^\mathrm{bg}$) for
background stars.  For both distributions the variance was set to the
sum of $\sigma^2_{A_V}$, the estimated variance in the measured column
density of the star, and $\sigma_{A_V}^\mathrm{fg2}$ (respectively,
$\sigma_{A_V}^\mathrm{bg2}$), the intrinsic variance of the foreground
(background) distributions.  The parameter $f \in [0, 1]$ used in
Eq.~\eqref{eq:6} represents the \textit{filling factor}, i.e.\ the
ratio between the area of the sky occupied by the cloud and the area
of the whole field.  

We considered this simple model leaving as free parameters $\bigl\{
\pi_\mathrm{Pipe}, f, A_V^\mathrm{bg}, \sigma_{A_V}^{\mathrm{bg}2}
\bigr\}$ (the average foreground $A_V$ and its variance were
deduced from the stars with measured parallax $\hat \pi > 12 \mbox{
  mas}$).  In order to assess the goodness of a model, we computed the
likelihood function, defined as
\begin{equation}
  \label{eq:7}
  \mathcal{L} = \prod_{n=1}^N \int_0^\infty p\bigl( A_V^{(n)} \bigm| \pi
  \bigr) p\bigl( \pi \bigm| \hat\pi^{(n)} \bigr) \, \diff \pi \; ,
\end{equation}
where the product is carried over all $N$ observed stars with
parallax, and where $p\bigl( \pi \bigm| \hat\pi^{(n)} \bigr)$ is the
probability distribution that the $n$-th star with measured parallax
$\hat\pi^{(n)}$ has a true parallax $\pi$.  The integral in
Eq.~\eqref{eq:7} takes into account the possibility that a star with a
measured parallax $\hat\pi < \pi_\mathrm{Pipe}$ (or $\hat\pi >
\pi_\mathrm{Pipe}$) is incorrectly taken as background (respectively,
foreground) because of its large parallax error $\hat\sigma_\pi$.  In
our analysis, we used for $p\bigl( \pi \bigm| \hat\pi^{(n)} \bigr)$ a
simple normal distribution with standard deviation equal to the formal
error on the measured parallax $\hat\sigma_{\pi}^{(n)}$:
\begin{equation}
  \label{eq:8}
  p\bigl( \hat\pi \bigm| \pi^{(n)} \bigr) = \mathrm{Gau}\bigl( \hat\pi
  \bigm| \pi^{(n)}, \hat\sigma_{\pi}^{(n)2} \bigr) \; .
\end{equation}
We note that this particular choice allows us to evaluate the
integrals of Eq.~\eqref{eq:7} analytically in terms of the error
function erf.

In order to study in detail the likelihood function \eqref{eq:7} in
its multidimensional parameter space we used Monte Carlo Markov Chains
(MCMC; see, e.g. \citealp{Tanner}).  The obtained results are
summarized by Fig.~\ref{fig:12}, where we show the marginalized
likelihood as a function of the cloud parallax $\pi$.  As best
estimate for the cloud distance we take the median of the distribution
shown in Fig.~\ref{fig:12}, $d_\mathrm{Pipe} = 130 \mbox{ pc}$; the
formal $68\%$ (respectively, $95\%$) confidence regions are $[110,
143] \mbox{ pc}$ and $[72, 154] \mbox{ pc}$.  Note that the best fit
value obtained for the filling factor is $f = 0.42$, which compares
well with the value directly measured in the map of Fig.~\ref{fig:7}
($41\%$ of the area in our field has a measured extinction $A_V > 3
\mbox{ mag}$).

In summary, in the rest of this paper we will use a estimate of the
cloud distance $130^{+13}_{-20} \mbox{ pc}$.  The formal error is
relatively small, but it is probably underestimated (e.g., the error
due to a possible misclassification of the spectral type of a star was
not included in the error budget); moreover, because of the relatively
small number of reddened stars, the distance obtained appears to be
slightly model-dependent.  Our distance estimate is likely to be
biased towards large values because the method used gives an
\textit{upper\/} limit (this is also suggested by the long tail at
large parallaxes in the likelihood function of Fig.~\ref{fig:12}).
With the advent of the new generation astrometric missions such as
Gaia \citep{1996A&AS..116..579L} it will be possible with this or
similar methods to accurately measure the distances of a large sample
of molecular cloud complexes.

\section{Statistical analysis}
\label{sec:statistical-analysis}

\subsection{Star extinction}
\label{sec:star-absorptions}

\begin{figure}[!tbp]
  \begin{center}
    \includegraphics[bb=141 298 453 531, width=\hsize]{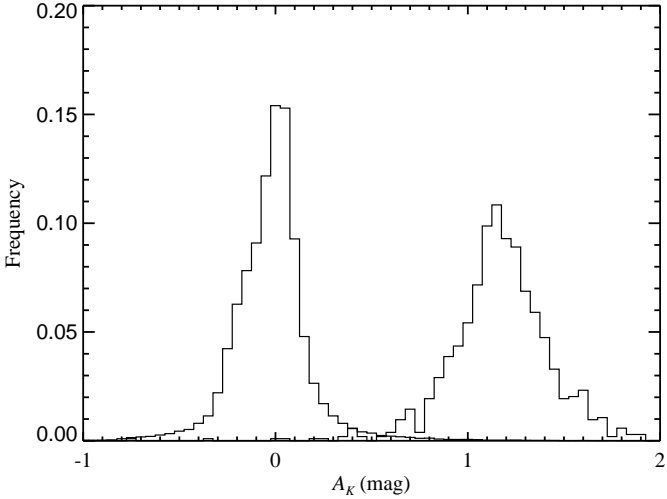}
    \caption{The distribution of individual star extinctions.  The
      solid histogram shows the distribution of column densities for
      stars observed in low-extinction regions ($-0.05 < A_K < 0.05$
      on the extinction map); the dashed line the same distribution
      for stars in high-extinction regions ($1.15 < A_K < 1.25$).}
    \label{fig:13}
  \end{center}
\end{figure}

In Fig.~\ref{fig:13} we plot the distributions of column densities
obtained for stars observed in low-extinction ($-0.05 \mbox{ mag} <
A_K < 0.05 \mbox{ mag}$) and high-extinction ($1.15 \mbox{ mag} < A_K
< 1.25 \mbox{ mag}$) regions.  The two histograms show two similar
Gaussian shapes, but the one corresponding to high-extinction has a
larger width (the best-fit Gaussian dispersions are $\sigma_1 = 0.210
\mbox{ mag}$ and $\sigma_2 = 0.272 \mbox{ mag}$).  This increase can
be due to a number of factors: (i) the increase of photometric errors
for the faint stars observed through dense clouds, and the lack of the
$K$ band photometry for most of these objects; (ii) the internal
structure of the cloud on scales smaller than our resolution (which,
we recall, is $\mathrm{FWHM} = 1 \mbox{ arcmin}$); (iii) an increase
in the contamination of foreground stars (see below
Sect.~\ref{sec:foregr-star-cont}).  Among these factors, (i) can be
evaluated from the photometric errors of the 2MASS database, and
appears to negligible (i.e., there is no significant increase in the
photometric error of stars in moderately extincted regions, cf.\
Fig.~\ref{fig:10}).  On the other hand, factors (ii) and (iii) are
difficult to disentangle, because they essentially produce the same
effects (see below).

\begin{figure}[!tbp]
  \begin{center}
    \includegraphics[bb=15 7 325 212, width=\hsize]{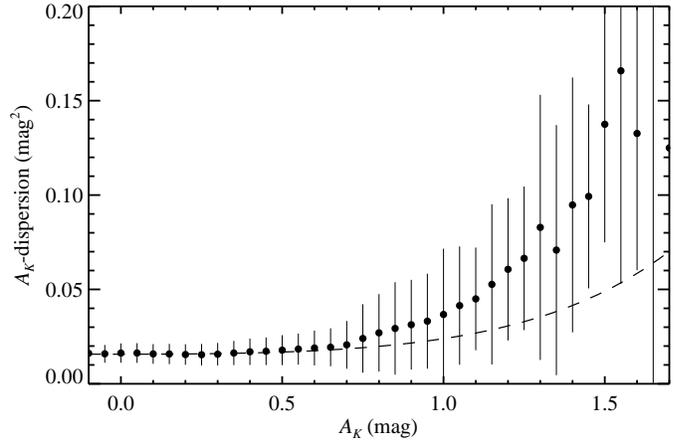}
    \caption{The dispersion on the extinction measurements as a
      function of the column density $A_K$.  In order to have a clear
      plot, we evaluated the dispersion on bins of $0.05$ magnitudes in
      $A_K$; the plot shows the average values obtained in each bin
      and the relative scatter.  The dashed line represents the
      expected increase in the dispersion if we attribute this effect
      to the contamination of foreground stars.}
    \label{fig:14}
  \end{center}
\end{figure}

The increase of the dispersion on $A_K$ at large column densities can
be evaluated more quantitatively from Fig.~\ref{fig:14}.  This figure
has been obtained from the map of the dispersion of $A_K$ (not shown
here).  This map is constructed by evaluating, for each pixel of
Fig.~\ref{fig:7}, the scatter of the column density estimates
corresponding to each star.  Hence, large values in some pixels of
this map mean that the stars used to estimate $A_K$ in the
corresponding pixels of Fig.~\ref{fig:7} present very different
reddening values (see Paper~I for the analytic definition of the
scatter map).  Figure~\ref{fig:14} has been obtained by plotting the
average values observed in the dispersion map in pixels characterized
by a given value of $A_K$ in Fig.~\ref{fig:7}.  We note that the
observed increase in the dispersion of $A_K$ has been observed in
several similar studies \citep[e.g.][]{1994ApJ...429..694L,
  1998ApJ...506..292A, 1999ApJ...512..250L}.  

As we mentioned above, in our case it is difficult to attribute the
effect observed in Fig.~\ref{fig:14} with certainty to foreground
stars or to substructures.  Consider, as an example, a
``two-dimensional Swiss cheese'' molecular cloud, i.e.\ a thin cloud
with many holes (some of them smaller than the resolution of our
maps).  In this case, a background star observed through a small hole
will produce exactly the same statistics as a foreground star.  Hence,
in this case it will be impossible to distinguish foreground stars and
substructure.  In less extreme (and more physical) situations it is
generally possible to firmly identify, in the dense regions of the
cloud complexes, foreground stars.  However, this task is
unfortunately non-trivial for the Pipe, because of its 
filamentary structure and of the relatively low resolution attainable
with the 2MASS data.
Still, our estimate indicates that only an extremely small fraction
$F_0 \simeq 0.003$ of foreground stars is present in our map (see
below Sect.~\ref{sec:foregr-star-cont}), and thus
they have a negligible contribution to the effect observed in
Fig.~\ref{fig:14} (cf.\ dashed line in that figure).  In summary, we
attribute most of the increase in the dispersion of $A_K$ to
unresolved substructures.  A further indication of this is given by
similar analyses of molecular clouds carried out at a higher
resolution \citep[e.g.][]{2004ApJ...610..303L}, which show a
dispersion $\sigma_{A_V}$ substantially independent of $A_K$ (i.e., a
flat curve in the $A_K$-$\sigma_{A_K}$ plot).  In a follow-up paper
\citep{L05} we will further investigate the effect of substructures in
molecular clouds.

\subsection{Reddening law}
\label{sec:reddening-law}

\begin{figure}[!tbp]
  \begin{center}
    \includegraphics[bb=152 299 457 531, width=\hsize]{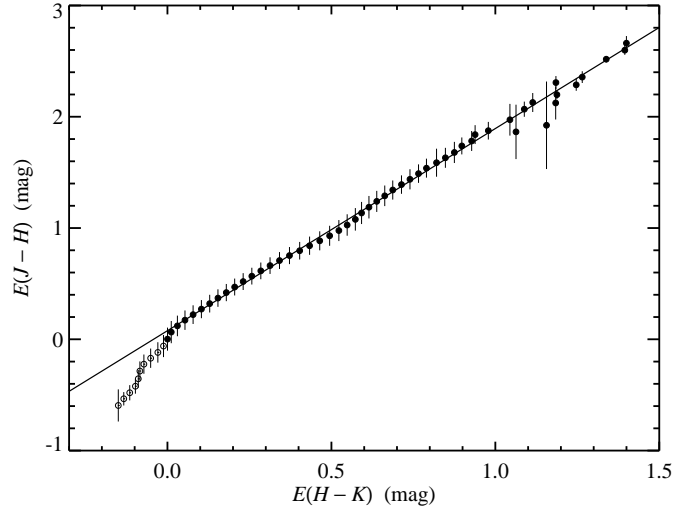}
    \caption{The reddening law as measured on the analyzed region.
      The plot shows the color excess on $J-K$ as a function of the
      color excess on $H - K$ (the intrinsic colors, deduced from the
      control field, are $0.18$ and $0.60$ respectively); the solid
      line shows the best fit.  Error bars represents the standard
      deviation of color excesses inside the bin.}
    \label{fig:15}
  \end{center}
\end{figure}

The reddening law plays a fundamental role in any infrared dust
measurement since it is used to derive the column dust density from
the properties of background stars.  In particular, a key assumption
is that the reddening law is \textit{linear}, i.e.\ that the ratio
$A_{\lambda_1} / A_{\lambda_2}$ of extinctions in two different
wavelengths is constant.  Since the intrinsic luminosities of stars
are not known, we cannot measure directly the extinction $A_\lambda$
in a single band; in the NIR, however, we can measure with good
accuracy the color excess $E(m_{\lambda_1} - m_{\lambda_2}) =
A_{\lambda_1} - A_{\lambda_2}$, and this is the essence of the
\textsc{Nice} and \textsc{Nicer} techniques.  Moreover, if we have
good photometry in three bands, say $J$, $H$, and $K$, we can
indirectly verify the assumption of a linear reddening law by checking
that the color excesses $E(H - K)$ and $E(J - H)$ are linearly
dependent.  The large number of stars in the field allows us to
perform this check with great accuracy.

The linearity of the reddening law, ultimately, is equivalent to say
that the stars in the color-color plot are mostly found along a linear
stripe, and the hypothesis is qualitatively well supported by
Fig.~\ref{fig:6}.  Still, this simple observation is not easily
translated into a \textit{robust\/} and \textit{statistically
  accurate\/} method to measure the reddening law.  This, in part,
might be the origin of some discrepancies found in the literature
about the slope of the reddening law (another good reason are the
different NIR photometric systems used by different authors).

For example, in order to measure the slope of the reddening law, we
can fit all star colors (or color excesses) with a linear relation:
\begin{equation}
  \label{eq:9}
  (J - H) = \alpha (H - K) + \beta \; ,
\end{equation}
where $\alpha = (A_J - A_H) / (A_H - A_K)$ and $\beta$ are taken to be
constant for all stars.  Naively, the fit could be done directly on
the individual star colors $\{H_n - K_n\}$ and $\{J_n - H_n\}$, by
minimizing the expression
\begin{equation}
  \label{eq:10}
  \chi^2 = \sum_{n=1}^N \bigl[ \alpha (H_n - K_n) + \beta - 
  (J_n - H_n) \bigr]^2 \; ,
\end{equation}
or similar expressions taking into account the photometric errors of
stars.  In reality, such a procedure leads to unsatisfactory results
because the large majority of stars are subject to small or negligible
reddening.  In other words, a minimization of Eq.~\eqref{eq:10} would
basically take into account only the lower-left part of the plot of
Fig.~\ref{fig:6}, and would neglect the most interesting (highly
reddened) stars (say, $H - K > 1 \mbox{ mag}$).

In order to equally take into account all reddening regimes, we followed
here a different procedure:
\begin{enumerate}
\item We evaluated the column density of each star by using the normal
  reddening law considered in this paper \citep{2005ApJ...619..931I}.
\item We binned the stars according to their extinction in 
  bins of $0.02 \mbox{ mag}$.
\item In each bin, we evaluated the average color excesses in $H - K$
  and $J - H$, and the standard deviation in $J - H$.  Note that the
  standard deviation in $J - H$ was calculated by referring all star
  colors inside the bin to the average $H - K$ color, or equivalently
  by evaluating the standard deviation of the expression
  \begin{align}
    \label{eq:11}
    E(J - H)_\mathrm{corr} = {} & -\alpha \bigl[ E(H - K) -
    \bigl\langle E(H - K) \bigr\rangle \bigr] \notag\\
    & {} + E(J - H) \; .
  \end{align}
  This procedure ensures that the scatter in $E(J - H)_\mathrm{corr}$
  does not include the scatter due to the finite size of the bin in $K
  - H$.
\item We fitted all \textit{positive\/} bins with a linear
  relationship of the form of Eq.~\eqref{eq:9} (in the fit we took
  into account the standard deviations evaluated in the previous point).
\item Finally, we repeated the whole procedure using the newly
  determined reddening law until convergence is achieved (which
  typically happens after two or three iterations).
\end{enumerate}
The method presented above has many pleasant properties.  First, all
reddening regimes equally contribute to the fit, since by construction
we equally use all bins of different column densities.  This, in turn,
implies that the method is robust against bright stellar populations
that might artificially change the slope of the reddening law (cf.\
above Sect.~\ref{sec:nicer-absorpt-map}).  Moreover, the use of the
\textit{observed\/} standard deviations on the $J - H$ color for the
fit properly takes into account local deviations from a reddening law
or increased photometric errors for reddened stars.

We applied the method described above to our data, by selecting all
2MASS stars with accurate photometry in all bands (we required all
photometric errors to be smaller than $0.1 \mbox{ mag}$) from the
\textit{cleaned\/} catalog (cf.\ above
Sect.~\ref{sec:nicer-absorpt-map}); the results obtained are
summarized in Fig.~\ref{fig:15}.  We found as best fit slope $\alpha =
1.82 \pm 0.03$, a value that appears to be in excellent agreement both
the \citet{2005ApJ...619..931I} 2MASS reddening law, $E(J - K) / E(H -
K) = 1.78 \pm 0.15$, and with the \citet{1985ApJ...288..618R} $E(J -
K) / E(H - K) = 1.70 $ normal reddening law converted into the 2MASS
internal photometric system \citep{2001AJ....121.2851C}, which is $E(J
- K)_\mathrm{2MASS} / E(H - K)_\mathrm{2MASS} \simeq 1.89$.

In follow-up papers we will apply the same technique to other cloud
complexes studied from the 2MASS archive.  The uniformity of the 2MASS
data and of the procedure used to derive the reddening law will allow
us to accurately study cloud-to-cloud variations (see, e.g.,
\citet{1998AJ....115..252K} for a case where significant differences
from a standard reddening law are found).

\subsection{Smoothing method}
\label{sec:smoothing-method}

\begin{figure}[!tbp]
  \begin{center}
    \includegraphics[width=\hsize]{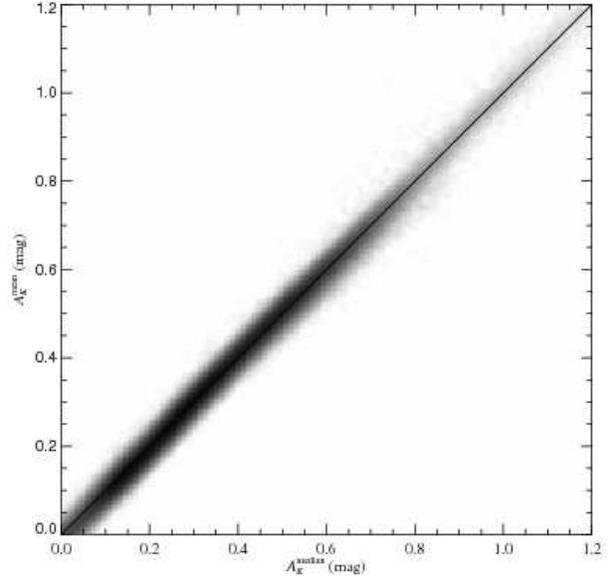}
    \caption{Plot of the column densities for the various pixels of
      the extinction map obtained through median filter and simple
      average.  The diagonal line shows the locus $A_K^\mathrm{median}
    = A_K^{mean}$.}
    \label{fig:16}
  \end{center}
\end{figure}

\begin{figure}[!tbp]
  \begin{center}
    \includegraphics[bb=140 313 457 518, width=\hsize]{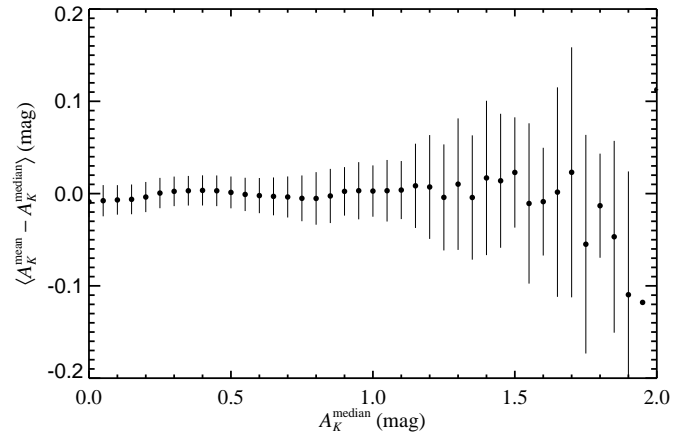}
    \caption{This plot, similarly to Fig.~\ref{fig:16}, shows the
      relationship between the pixel column densities measured using
      the median and the simple average.  The plot has been produced
      by splitting the pixels in bins of $0.05$ magnitudes in
      $A_K^\mathrm{median}$, and by plotting, for each bin, the
      average and the 2-$\sigma$ scatter of the quantity
      $A_K^\mathrm{mean} - A_K^\mathrm{median}$.}
    \label{fig:17}
  \end{center}
\end{figure}

\begin{figure}[!tbp]
  \begin{center}
    \includegraphics[bb=135 313 457 518, width=\hsize]{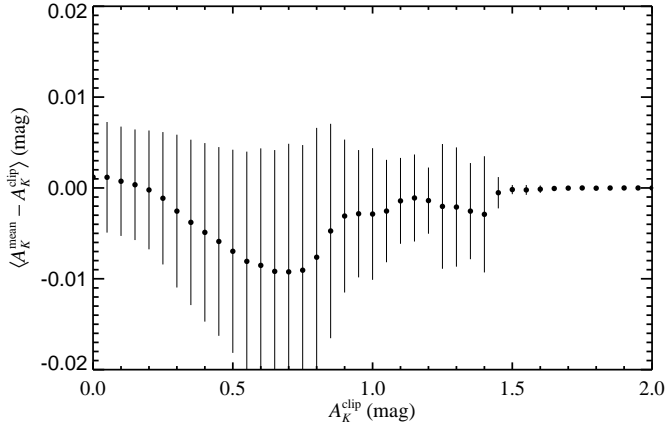}
    \caption{This figure compares the column densities obtained
      through sigma-clipping and simple average, similarly to what
      done in Fig.~\ref{fig:17} for the median.}
    \label{fig:18}
  \end{center}
\end{figure}

As discussed in Paper~I, the pipeline developed allows us to obtain
smooth maps of the discrete column density measurements (obtained for
each background star) in three different ways: simple average,
sigma-clipping, and median map.  For the simple average we use all
column density measurements close to each pixel, and the extinction
value assigned to the pixel is a weighted average of the $A_K$
measurements for each star.  The weight takes into account both the
statistical error on $A_K$ (which is due to the photometric errors and
to the intrinsic scatter of star colors), and the angular distance
between the star and the pixel.  For the sigma-clipping, instead, we
selectively discard outliers on $A_K$, i.e.\ stars that have
associated column densities significantly different from the local
average.  Finally, the median map is constructed by evaluating a sort
of ``weighted median'' for the column densities measured for angularly
close stars.

Figure~\ref{fig:16} shows the relationship between the extinction
measurements obtained through simple average and median.  The very
tight band show that the two methods give comparable results.  A more
quantitative analysis is provided by Fig.~\ref{fig:17}, where an
histogram of the differences between the simple mean and the median
for each pixel is presented.  Note that the two methods are
statistically indistinguishable up to $A_K \simeq 1.5 \mbox{ mag}$,
where the mean starts to estimate consistently smaller column
densities.  Hence, this confirms that the use of the simple mean is
justified up to relatively high column densities.  Figure~\ref{fig:17}
also shows an increase in the scatter at high column densities, an
indication of a difference in the extinction estimates at the
corresponding sky locations using the mean and the median smoothing
techniques.  Unresolved structures in the cloud, which are expected in
the dense regions, are the most probably reason for the observed
scatter (substructures are likely to play different roles in the
simple mean and median smoothing).

A statistical comparison between the simple mean and the
$\sigma$-clipping smoothing techniques is shown in Fig.~\ref{fig:18}.
From this figure, we can deduce that there is no significant
difference in using these two techniques.  This is mostly due to the
fact that the two methods are expected to provide different results
only at high $A_K$, when a possibly significant number of foreground
stars might contaminate the map.  On the other hand, these regions
also have a much smaller star density, and thus the extinction map
there will have a large error (see Fig.~\ref{fig:8}).  This, in turn,
implies that the sigma clipping is not effective there (the
statistical error on the map is large enough to accommodate, within a
3-$\sigma$ interval, unreddened stars).  Hence, in the following we
will only consider the median and the simple mean estimator.

In conclusion, our analysis shows that the simple mean estimator,
which has the smallest scatter, is reliable up to approximately $A_K
\simeq 1.5 \mbox{ mag}$; for larger column densities, the median
should be used in order to minimize the effect of foreground stars.

\subsection{Foreground star contamination}
\label{sec:foregr-star-cont}

\begin{figure}[!tbp]
  \begin{center}
    \includegraphics[bb=167 301 445 522, width=\hsize]{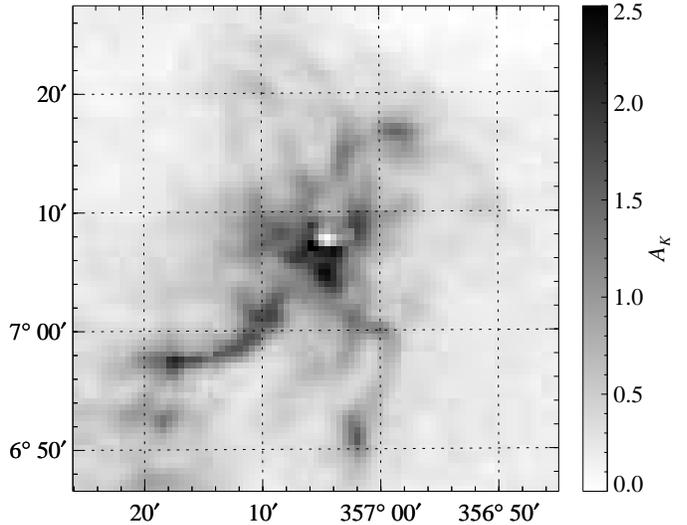}
    \caption{A zoom of the extinction map showing Barnard~59.
      Note the white ``hole'' close to the center of this cloud, which
      is due to the combined effect of very high extinction values and
      the presence of embedded stars in this star forming core.}
    \label{fig:19}
  \end{center}
\end{figure}

The three smoothing methods described above differ significantly on
the effect of foreground contamination.  If we assume that a fraction
$F > 0$ of stars are foreground with respect to the cloud, then the
simple average estimate will be biased toward small column densities.
In particular, the average measurement of $A_K$ will be $\langle \hat
A_K \rangle = (1 - F) A_K^\mathrm{true}$.  In contrast, the median
will provide an (almost) unbiased estimate of $A_K$ as long as $F <
0.5$ (see \citealp{2002AJ....123.2559C} and \citealp{Lombardi2004} for
a detailed discussion of the median properties).  The sigma-clipping
estimate will often be between these two extremes: it will be
effective in removing foreground stars only in relatively dense
regions and only for small values of $F$.

Because of selection effects, the value of $F$ changes significantly
on the field, and in particular increases in high-column density
regions.  This, in turn, implies that the correction to be used on the
estimated value of $A_K$ is not constant on the field.  In the case of
the Pipe nebula only a few magnitude extinction in $V$ is observed in
most of the field (see above Sect.~\ref{sec:nicer-absorpt-map}; see
also Fig.~\ref{fig:26} below), and thus the value of $F$ is expected
to be approximately constant.  Moreover, since this cloud complex is
very close to us and is observed close to the galactic center, we
expect only a tiny fraction of foreground stars.

In order to evaluate quantitatively the density of foreground stars,
we have selected high-extinction regions characterized by $A_K > 0.8
\mbox{ mag}$ and with relatively small expected error in $A_K$ (we
allowed for a maximum error of $0.2 \mbox{ mag}$).  In order to avoid
the effects of possible substructures and ambiguities in the
identification of foreground stars, we restricted our analysis to the
main core of the Pipe nebula, i.e. to the region at $l \simeq
1.5^\circ$ and $b \simeq 4^\circ$ (see Fig.~\ref{fig:7}).  We have
then checked all stars in these regions that show ``anomalous''
extinction, i.e.\ stars whose column densities differ by more than
3-$\sigma$ with respect to the field.  A total of 70 stars met this
criteria; hence, since the area selected is approximately $1057 \mbox{
  arcmin}^2$, we estimated a foreground star density of $0.066 \mbox{
  arcmin}^{-2}$.  As a result, the relative fraction of foreground
stars in regions with negligible absorption is only $F|_{A_V=0} \equiv
F_0 = 0.003$, and we can safely ignore the effect of foreground stars
except on the higher extinction regions.  Note, in particular, that a
clear sign of very high extinction (very few background stars) and
contamination by foreground stars is observed only in Barnard~59,
where we see a ``hole'' in the extinction close to the center of this
clump (see Fig.~\ref{fig:19}).  In this case, these foreground stars
are not really foreground but young stars moderately embedded in
Barnard~59, an active star forming region
\citep[e.g.][]{1999PASJ...51..871O}.  Our estimate of $F_0$ allows us
to evaluate the maximum theoretical extinction measurable with the
\textsc{Nice} and \textsc{Nicer} method: using Eq.~(53) of
\citet{Lombardi2004}, we obtain $A_K^\mathrm{max} \simeq 2.9 \mbox{
  mag}$, which is close to the maximum value measured here.

\subsection{Column density distribution}
\label{sec:pixel-distribution}

\begin{figure}[!tbp]
  \begin{center}
    \includegraphics[bb=141 327 451 504, width=\hsize]{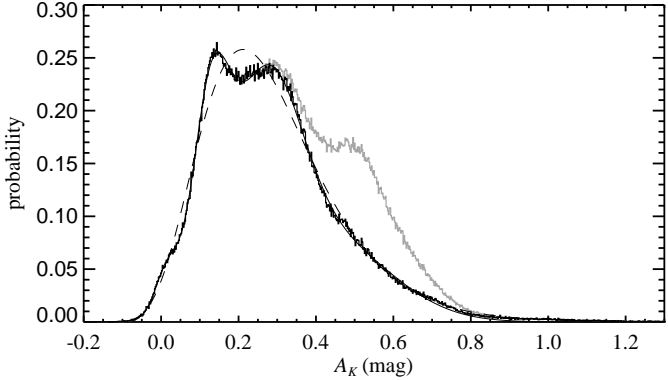}
    \caption{The distribution of pixel extinction over the whole
      field.  The figure shows the histogram of column densities for
      all pixels (gray line) and for the pixels with galactic latitude
      $b > 3^\circ$.  Note that the latter histogram can be fitted
      almost perfectly using four normal distributions (solid line) or
      four lognormal distributions (not shown here), while a single
      lognormal (dashed lined) is able to reproduce the general shape
      of the histogram.}
    \label{fig:20}
  \end{center}
\end{figure}

\begin{table}[b!]
  \centering
  \begin{tabular}{lcccc}
    Distr.    & $x_0$    & $x_1$   & $\sigma$& rel. factor \\
              & (mag)    & (mag)   & (mag)   &             \\
    \hline
    \hline
    lnGau \#1 & $-0.241$ & $0.503$ & $0.325$ & $1.000$ \\     
   \hline
    lnGau \#1 & $-0.150$ & $0.413$ & $0.501$ & $0.150$ \\
    lnGau \#2 & $-1.805$ & $1.943$ & $0.018$ & $0.744$ \\
    lnGau \#3 & $0.039$  & $0.267$ & $0.371$ & $0.106$ \\ 
    \hline
    Gau \#1   & $0.015$  & --      & $0.022$ & $0.086$ \\
    Gau \#2   & $0.123$  & --      & $0.033$ & $0.353$ \\ 
    Gau \#3   & $0.267$  & --      & $0.064$ & $0.395$ \\
    Gau \#4   & $0.419$  & --      & $0.118$ & $0.166$
  \end{tabular}
  \caption{The best-fit parameters obtained from a fit of the column
    density distribution shown in Fig.~\ref{fig:20} using a single
    lognormal distribution, three lognormal distributions, or four
    normal distributions.}
  \label{tab:1}
\end{table}

A fundamental statistical property of a cloud is the distribution of
column densities.  \citet{1994ApJ...423..681V} showed that for highly
supersonic flows (such that the Mach number $M = u / c_\mathrm{s} \gg
1$), the gas essentially has a pressureless behavior, and
gravitational forces can also become negligible.  In these conditions,
the hydrodynamic equations become scale-invariant: in other words,
motions at different length and density scales obey essentially the
same equations.  For a fully developed turbulent gas, density and
velocity can be regarded as random fields.  Because of the scale
invariance, the probability of having a local (volume) density
fluctuation of amplitude $\Delta \rho$ depends uniquely on the ratio
$\Delta \rho / \rho$, i.e.\ on the \textit{relative\/} fluctuation
amplitude; hence, the probability distribution function of the density
at each point of the cloud is expected to be lognormal.  When
projecting the three-dimensional mass density along the line of sight,
the lognormal distribution is essentially preserved, i.e.\ the
projected two-dimensional density is also expected to be well
approximated by a lognormal distribution.

More recently, \citeauthor{1994ApJ...423..681V} argument has been
investigated in more detail using various simulations by a number of
authors \citep[e.g.][]{1997ApJ...474..730P, 1997MNRAS.285..711P,
  1997ApJ...481L..27P, 1998PhRvE..58.4501P, 2000ApJ...535..869K}, and
in all cases the lognormal distribution was found to be a good
approximation of the (projected) cloud density.  However,
\citet{1998ApJ...504..835S} found that, contrary to previous claims,
the density probability distribution is well approximated in their
simulations by an exponential law in a relatively large range of
physical parameters.  Recently, this argument has been challenged by
\citet{2001ApJ...546..980O} by showing a good agreement with a
lognormal distribution in the molecular cloud IC5146
\citep{1999ApJ...512..250L}.

A direct study of the extinction distribution on a real cloud is a
stringent test for simulations and ultimately also provides important
hints on the physical conditions inside molecular clouds.  We
evaluated the column density distribution in the Pipe nebula by
constructing the histogram of extinction measurements in each pixels
of the map shown in Fig.~\ref{fig:7} with $b > 3^\circ$, binned on
suitable intervals (we used here $0.002 \mbox{ mag}$ in $A_K$).  The
obtained distribution, plotted in Fig.~\ref{fig:20}, shows a complex
shape with multiple peaks.  

In order to compare the empirical distribution with a theoretical one,
we used Poisson models for the individual histogram bins, and
evaluated the joint probability to have the observed histogram in the
range $-0.3 \mbox{ mag} < A_V < 2 \mbox{ mag}$ (we excluded high $A_K$
bins in order to avoid the complications inherent with large-column
density regions).  We constructed the theoretical models as sum of
normal distributions
\begin{equation}
  \label{eq:12}
  \mathrm{Gau}(x | x_0, \sigma^2) = \frac{1}{\sqrt{2 \pi \sigma^2}} \exp
  \left( -\frac{(x - x_0)^2}{2 \sigma^2} \right) \; ,
\end{equation}
and lognormal ones
\begin{align}
  \label{eq:13}
  &\mathrm{lnGau}(x | x_0, x_1, \sigma) = {} \notag\\
  & \quad {} = \frac{1}{(x - x_0) \sqrt{2
      \pi \sigma^2}} \exp \left( -\frac{\bigl[\ln (x - x_0) - \ln x_1
      \bigr]^2}{2 \sigma^2} \right) \; .
\end{align}
As shown by Fig.~\ref{fig:20}, the best fit obtained from a single
lognormal (dashed line) is surprisingly good given the complexity of
the distribution; clearly, however, in this case we are only able to
reproduce the general shape of the histogram and we are unable to
generate finer details clearly visible in Fig.~\ref{fig:20}.  As a
measure of the quality of the fit, we note that the residuals are more
than $150 \sigma$ away from the observed histogram.

In order to improve the fit we added more components.  However,
two-component distributions still gave unsatisfactory results, and we
had to use more components distributions to obtain a close match to
the data.  In particular, we verified that both the sum of four normal
and three lognormal distributions gave similar fits, with residuals of
the order of $8 \sigma$.  Hence although, in principle, the fit is
still statistically inconsistent with the data, we note that in
practice the large number of bins and the non-trivial structure of the
Pipe nebula make it difficult to obtain good fits with a relatively
small number of parameters.  Note also that the need for several
components in the column-density distribution is reminiscent of the
velocity structures observed in the Pipe nebula from CO data
\citep{1999PASJ...51..871O}.  In this respect, it is also likely that
some of the ``weak'' components (e.g., the lnGau~\#3 or the Gau~\#4) are
related to background clouds observed in projection to the Pipe nebula
at low galactic latitudes.

\section{Comparison with CO analysis}
\label{sec:comparison-with-co}

\begin{figure*}[t]
  \centering
  \includegraphics[bb=148 303 434 522, width=0.48\hsize]{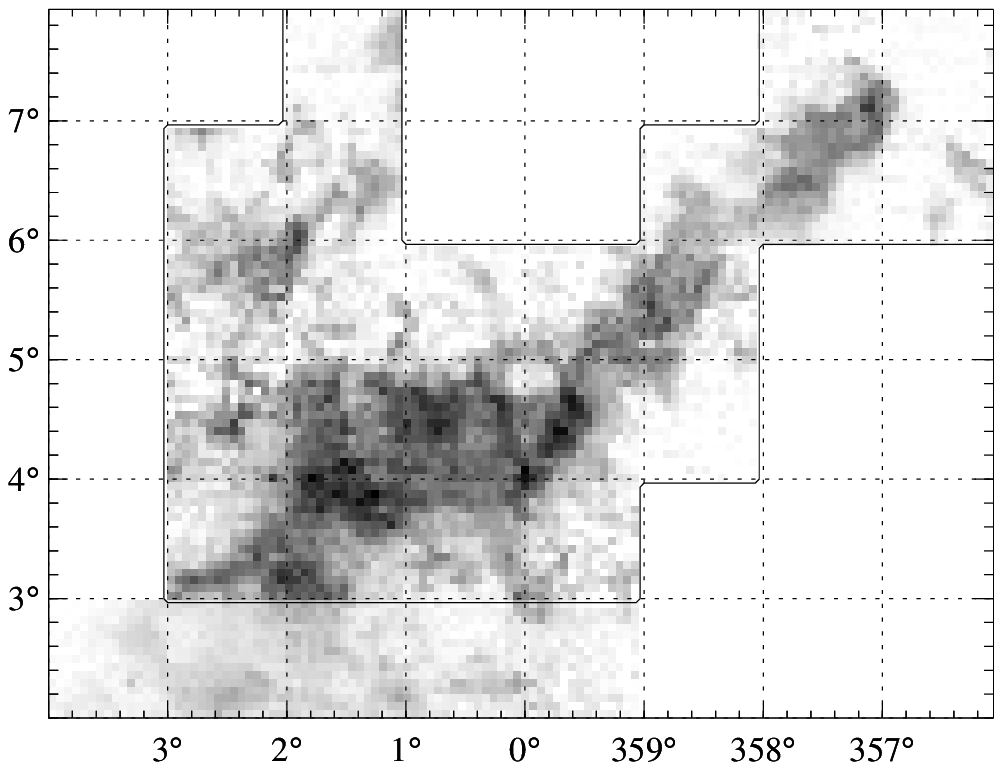}
  \includegraphics[bb=148 303 434 522, width=0.48\hsize]{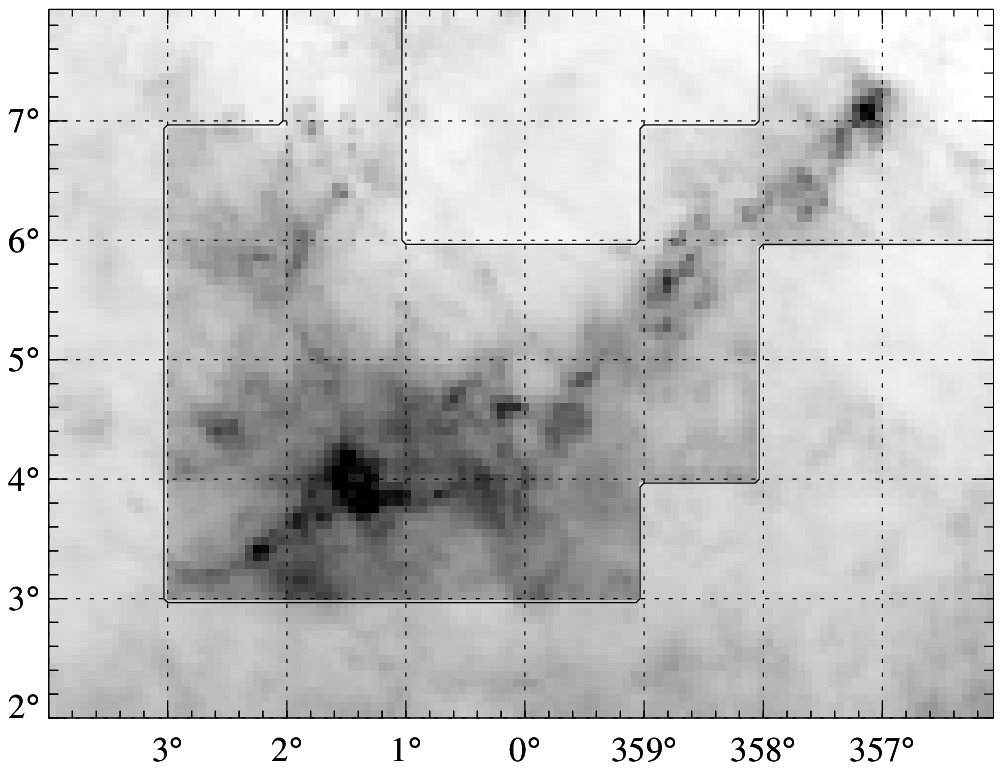}
  \caption{Left: The NANTEN integrated $^{12}$CO column density map
    (kindly provided by \citealp{1999PASJ...51..871O}); the white
    regions have not been observed and no data are thus available
    there; the shaded region is located at $b < 3^\circ$ and has been
    excluded from the analysis to avoid contamination from
    low-galactic latitude clouds.  Right: The \textsc{Nicer}
    extinction map downgraded to the resolution of the NANTEN map;
    shaded regions are excluded from the analysis.}
  \label{fig:21}
\end{figure*}

Radio observations of H${}_2$ surrogates, and in particular of CO
isotopes, provide an alternative independent estimate of the cloud gas
column density.  \citet{1999PASJ...51..871O} studied in detail the
Pipe nebula with the NANTEN radiotelescope and kindly provided their
${}^{12}$CO map to perform a comparison with the \textsc{Nicer}
analysis described in this paper.  To this purpose, we constructed an
extinction map using the same resolution (4~arcmin) and coordinate
system of the NANTEN observations.  In order to avoid any
contamination by low-galactic latitude clouds, we excluded all
measurements at $b < 3^\circ$ (see Fig.~\ref{fig:21}).

\begin{figure}[t]
  \centering
  \includegraphics[width=\hsize]{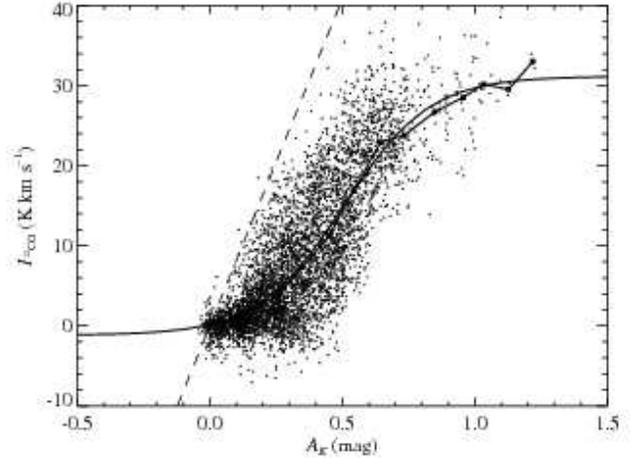}
  \caption{The $^{12}$CO integrated intensity $I_{{}^{12}\mathrm{CO}}$
    as a function of the \textsc{Nicer} extinction.  Each point of
    this plot shows the $^{12}$CO radio measurement from
    \citet{1999PASJ...51..871O} and the \textsc{Nicer} extinction at
    the corresponding position.  The filled squares represent the
    average $^{12}$CO intensity in bins of $0.1 \mbox{ mag}$ of
    $K$-band extinction; the solid, smooth line is the best fit for
    functions of the form of Eq.~\eqref{eq:14}; the dashed line shows
    the linear relation used by \citet{1999PASJ...51..871O} (X-factor
    $2.8 \times 10^{20} \mbox{ cm}^{-2} \mbox{ K}^{-1} \mbox{ km}^{-1}
    \mbox{ s}$).}
  \label{fig:22}
\end{figure}

Figure~\ref{fig:22} shows the relationship between the \textsc{Nicer}
column density and the NANTEN integrated $^{12}$CO temperature,
$I_{{}^{12}\mathrm{CO}} \equiv \int T \, \diff v$.  The large number
of independent measurements shown in this figure ($\sim 5\,000$) was
used to compare these two estimates of the gas column density and to
show the limits and merits of both.

\begin{figure}[t]
  \centering
  \includegraphics[width=\hsize]{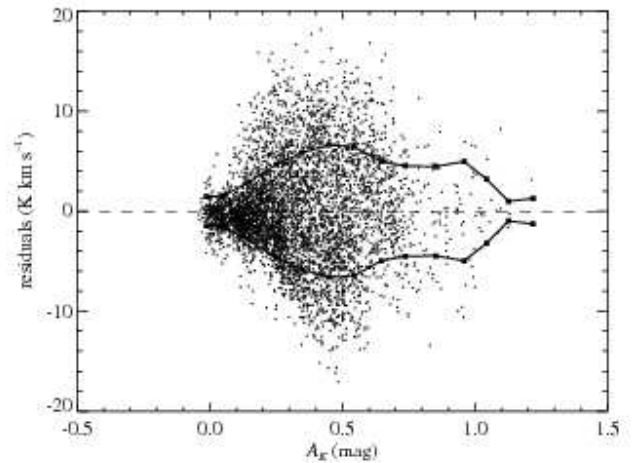}
  \caption{The residuals in the best-fit \eqref{eq:14} of
    Fig.~\ref{fig:22}.  The filled squares represent the standard
    deviations in bins of $0.1 \mbox{ mag}$.}
  \label{fig:23}
\end{figure}

From Fig.~\ref{fig:22} we can deduce a number of qualitative points.
We first note that apparently $^{12}$CO measurements are insensitive
to low column density regions.  Specifically, it appears that up to
$\sim 0.2 \mbox{ mag}$ of $K$-band extinction the radio measurements
are uniformly distributed around $0 \mbox{ K km s}^{-1}$. The
relatively small scatter observed in Fig.~\ref{fig:22} up to
$0.15$--$0.2 \mbox{ mag}$ of extinction (cf.\ also Fig.~\ref{fig:23})
indicates that both methods considered here have small intrinsic
internal errors.

At higher extinctions, and up to $A_K \simeq 0.6 \mbox{ mag}$, we
observe an almost linear relationship between the extinction and the
$^{12}$CO measurements; correspondingly, the scatter in the plot
increases significantly (see below).  We note that the presence of a
linear relationship between the CO integrated temperature and the NIR
extinction (and thus the hydrogen projected density), although on a
relatively small extinction range, is not obvious on the scales
considered here and for non-virialized cloud systems
\citep[see][]{1986ApJ...309..326D}.  Finally, for $A_K \gtrsim 0.7
\mbox{ mag}$, the $^{12}$CO data appear to saturate to a constant
value close to $30 \mbox{ K km s}^{-1}$.  This well-known saturation
effect is described in terms of an exponential relation between the
integrated temperature $I_{{}^{12}\mathrm{CO}} \propto 1 -
\mathrm{e}^{-\tau}$ and the cloud optical depth $\tau \propto A_K$.
The following analysis will mostly focus on the unsaturated $^{12}$CO
regime and will thus directly make use of the integrated temperature.

A more quantitative analysis of Fig.~\ref{fig:22} was carried out as
follows.  We divided the measurements in regular bins in $A_K$ (we
used a bin size of $0.1 \mbox{ mag}$), and we computed in each bin the
average of the $^{12}$CO intensity.  The results obtained are shown as
filled squares in Fig.~\ref{fig:22}.  This simple plot confirmed the
qualitative remarks discussed above and suggested that we could
approximate the $A_V$-$^{12}$CO relationship with a function of the
form
\begin{equation}
  \label{eq:14}
  I_{{}^{12}\mathrm{CO}} = A \left[ \frac{1}{1 + \exp \bigl[ - (A_K -
    A_K^\mathrm{mid}) k \bigr]} - b \right] \; ,
\end{equation}
We fitted this equation to the data by minimizing the scatters between
the predicted CO integrated intensity and the observed one; the best
fit parameters obtained were $A = 32.3 \mbox{ K km s}^{-1}$,
$A_K^\mathrm{mid} = 0.51 \mbox{ mag}$, $k = 6.20 \mbox{ K km s}^{-1}
\mbox{ mag}^{-1}$, and $b = 0.036$.  The residuals of this fit with
the data are shown in details in Fig.~\ref{fig:23}; the increase of
the dispersion in the relation \eqref{eq:14} at $A_K \simeq 0.2 \mbox{
  mag}$ is evident from this plot.  Since the expected error in the
\textsc{Nicer} map of Fig.~\ref{fig:21} is as low as $\sim 0.01 \mbox{
  mag}$, and since the expected error in the $^{12}$CO integrated
velocities is also relatively small (this can be estimated from the
residuals at $A_K \simeq 0 \mbox{ mag}$ of Fig.~\ref{fig:23}, and is
of order of $1.5 \mbox{ K km s}^{-1}$), we can deduce that the scatter
shown in Fig.~\ref{fig:23} for $A_K > 0.2 \mbox{ mag}$ is physical:
the ratio of dust and $^{12}$CO in the Pipe (and likely in other
molecular clouds) is far from being constant.

\begin{figure}[t]
  \centering
  \includegraphics[width=\hsize]{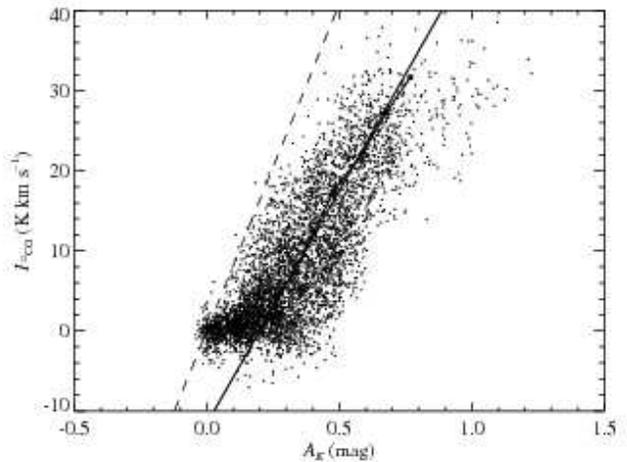}
  \caption{The $^{12}$CO-$A_K$ relation, with datapoints binned along
    the CO axis every $5 \mbox{ K km s}^{-1}$ (filled squares).  The
    solid line represents the best fit (on the whole field) from
    Eq.~\eqref{eq:15}.}
  \label{fig:24}
\end{figure}

\begin{figure}[t]
  \centering
  \includegraphics[width=\hsize]{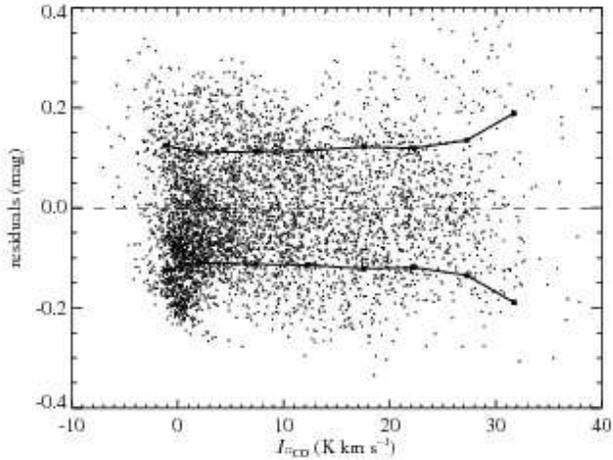}
  \caption{The scatter of our measurements on the linear fit of
    Eq.~\eqref{eq:15}.  Note how the derived standard deviation (filled
    squares) is practically constant over the CO column densities
    investigated here.}
  \label{fig:25}
\end{figure}

So far we investigated the $A_K$-$^{12}$CO relationship using the
value of $A_K$ as independent quantity: in other words, we studied the
expected CO measurement for each given $A_K$ column density.  We now
swap the role of $A_K$ and CO, and consider the average $A_K$ value
corresponding to a given $^{12}$CO measurement.  To this purpose, we
averaged the values of the \textsc{Nicer} extinction in bins of $5
\mbox{ K km s}^{-1}$.  The result, shown in Fig.~\ref{fig:24},
suggests that we can well approximate the average with a linear
relationship of the form
\begin{equation}
  \label{eq:15}
  A_K = A_K^{(0)} + r I_{{}^{12}\mathrm{CO}} \; .
\end{equation}
Note that we need to include explicitly a non-vanishing ``zero point'',
$A_K^{(0)}$, for the $A_K$ measurement. This is due the dissociation
of the CO molecule by the interstellar UV radiation field.  Our
results indicate that CO molecules in the Pipe become (self-) shielded
from the interstellar radiation field at about 1 magnitude of visual
extinction (2 magnitudes along the entire line of sight through the
cloud), consistent with standard theoretical predictions and prior
observations (e.g., \citealp{1988ApJ...334..771V},
\citealp{1999ApJ...515..265A}, \citealp{2002ApJ...570L.101B}). This CO
threshold should in principle be a function of the intensity of the
local interstellar radiation field and could in principle vary from
cloud to cloud.  We stress that it is highly unlikely that the
\textsc{Nicer} technique overestimates the extinction at low $A_K$,
i.e.\ that the ``zero point'' observed in the relation \eqref{eq:15}
is an artifact; rather, if there is a bias in \textsc{Nicer}, this is
likely to be toward an \textit{underestimate\/} of the column density
(because of a possible reddened control field).

Our data, combined with the $^{12}$CO data, allows for the best
determination of the CO-to-H$_2$ conversion factor (X-factor) using
dust as a tracer of H$_2$, because of the large number of measurements
(approximately $5\,000$) and also because our (smoothed) dust extinction
measurements have a mean error smaller than 0.05 magnitudes of visual
extinction.  To derive the X-factor for the $^{12}$CO data we
performed several best fits using Eq.~\eqref{eq:15} using different
selections of the points of Fig.~\ref{fig:24}; the results obtained
are reported in Table~\ref{tab:2}.

Finally, Fig.~\ref{fig:25} shows the residuals obtained from the fit
of Eq.~\eqref{eq:15}.  Interestingly, these residuals appear to be
rather constant on the field, and their standard deviation is
approximately $0.1 \mbox{ mag}$ in $K$-band extinction (or about 1
magnitude of visual extinction): hence, a single pointing in $^{12}$CO
can only constrain the dust column density with an accuracy of a few
magnitudes. Remarkably, this result is consistent with that found by
\citet{1994ApJ...429..694L} for ${}^{13}$CO in the molecular cloud
IC~5146, \citet{1999ApJ...515..265A} for ${}^{18}$CO in L977, and
\citet{2002ApJ...570L.101B} for the molecular cloud Barnard~68.
Hence, we find what it seems to be an irreducible uncertainty in the
dust-CO correlation of about 1 magnitude of visual extinction.  The
cause for this irreducible uncertainty is not known and it deserves a
dedicated study. A possible origin for this uncertainty might lie in
the sensitivity of the CO molecule (and its isotopes) to variations in
temperature (even along the same line-of-sight), density and UV
radiation field
\citep[e.g.][]{1982ApJ...262..590F,1988ApJ...334..771V}.  We note
however that changes in the abundance of $^{12}$CO, due to freezing
out on grains, are probably not in place at the extinctions less than $A_K
\simeq 1 \mbox{ mag}$ considered here when comparing the NANTEN and
2MASS/\textsc{Nicer} analyses.

\begin{figure}[!tbp]
  \begin{center}
    \includegraphics[bb=148 303 487 522,width=\hsize]{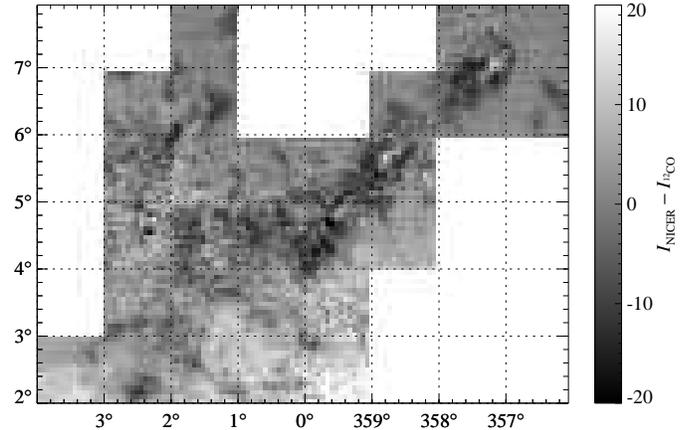}
    \caption{The difference between the two maps of Fig.~\ref{fig:21},
      i.e.\ the \textsc{Nicer} extinction map, converted into
      intensity using Eq.~\eqref{eq:14}, and the ${}^{12}$CO
      intensity.}
    \label{fig:30}
  \end{center}
\end{figure}

In order to further test this point, we considered the difference
between the \textsc{Nicer} extinction and the ${}^{12}$CO integrated
intensity.  In particular, we converted the degraded $A_K$
\textsc{Nicer} extinction map shown in Fig.~\ref{fig:21} (right) into
a ${}^{12}$CO intensity using Eq.~\eqref{eq:14}, and subtracted from
this map the NANTEN ${}^{12}$CO extinction map.  The result obtained,
shown in Fig.~\ref{fig:30}, gives insight into the origin of the
scatter observed in Fig.~\ref{fig:22}.  The first striking fact is
that clear structures are observed in Fig.~\ref{fig:30}, and this
alone rules out the possibility that the difference between the
2MASS-\textsc{Nicer} and the NANTEN measurements is due to statistical
errors.  Instead, in large areas we observe systematically positive or
negative values of the difference between the two maps.

In analysing Fig.~\ref{fig:30} one should always keep in mind that
this map has been built by converting the \textsc{Nicer} column
density into a radio intensity by using the best fit provided by
Eq.~\eqref{eq:14}; hence, differences have to be interpreted as
deviations from the fit used.  With this point in mind, we can deduce
some interesting facts from Fig.~\ref{fig:30}:
\begin{itemize}
\item The outer regions of the nebula appear to have consistent
  measurements in radio and NIR.  Given the form of Eq.~\eqref{eq:14},
  this further confirms that ${}^{12}$CO observations are insensitive
  to low column densities.
\item In many peripheral parts of the Pipe nebula, the ${}^{12}$CO
  intensity appears to overestimate the (converted) \textsc{Nicer}
  column density (see, e.g., the dark region in the middle of
  Fig.~\ref{fig:30}).  This effect, in principle, could be related to
  differences in the photo-dissociation of CO molecules due to
  variations of the local interstellar radiations field.  However,
  since generally an \textit{excess\/} of ${}^{12}$CO intensity is
  observed in the \textit{periphery\/} of the Pipe, it is more
  likely that the effect is related to temperature gradients in the
  nebula (this is also suggested by the patchy aspect of the map of
  Fig.~\ref{fig:30}).
\item The dense cores (e.g., Barnard~59) generally appear bright in
  Fig.~\ref{fig:30}, indicating that the ${}^{12}$CO map
  underestimates the column density there when compared to the
  \textsc{Nicer} analysis.  This effect is clearly due to the
  saturation effect of ${}^{12}$CO at high column densities (the
  saturation hence appears to be even stronger than suggested by
  Fig.~\ref{fig:22} and by the fit \eqref{eq:14}).
\end{itemize}
The items above proves that a joint analysis of the \textsc{Nicer}
and CO results can provide nontrivial and interesting insights into
the structure of molecular clouds and the process that regulate the
radio emission.  Clearly, the results obtained here are specific to
the Pipe nebula; however, by carrying out similar discussions for other
cloud complexes we hope to be able to draw general conclusions on this
important topic.

\begin{table}
  \centering
  \begin{tabular}{lccc}
    Fit & $A_K^{(0)}$ & $r$ \\
        & (mag)    & ($\mbox{mag K}^{-1} \mbox{ km}^{-1} \mbox{ s}$) \\
    \hline
    Whole sample& 0.176 & 0.0184 \\
    $I_{{}^{12}\mathrm{CO}} > 5 \mbox{ K km s}^{-1}$ & 0.199
    & 0.0172 \\ 
    $A_K > 0.1 \mbox{ mag}$ & 0.211 & 0.0165 \\
    $A_K \in [0.1, 0.6] \mbox{ mag}$ & 0.226 & 0.0127 \\
  \end{tabular}
  \caption{Best fit parameters relative to Eq.~\eqref{eq:15}.  If a
    normal reddening law is assumed, the X-factors derived from these
    fits are $\{ 4.21, 3.93, 3.77, 2.91 \} \times 10^{20} \mbox{ cm}^{-2}
    \mbox{ K}^{-1} \mbox{ km}^{-1} \mbox{ s}$ respectively.  The
    formal, $1\sigma$ errors are $0.002 \mbox{ mag}$ and $0.0002$ for
    $A_K^{(0)}$ and $r$, respectively (for all four fits).}
  \label{tab:2}
\end{table}

\section{Mass estimate}
\label{sec:mass-estimate}

The cloud mass $M$ can be derived from the $A_K$ extinction map using the
following simple relation
\begin{equation}
  \label{eq:16}
  M = d^2 \mu \beta_K \int_\Omega A_K \, \diff^2 x \; ,
\end{equation}
where $d = 130 \mbox{ pc}$ is the cloud distance, $\mu$ is the mean
molecular weight corrected for the helium abundance, $\beta_K \simeq
1.67 \times 10^{22} \mbox{ cm}^{-2} \mbox{ mag}^{-1}$ is the ratio
$N(\mathrm{H\textsc{i}}) + N(\mathrm{H}_2) / A_K$
(\citealp{1979ARA&A..17...73S}; see also \citealp{1955ApJ...121..559L,
  1978ApJ...224..132B}), and the integral is evaluated over the whole
field $\Omega$.  Assuming a standard cloud composition ($63\%$
hydrogen, $36\%$ helium, and $1\%$ dust), we find $\mu = 1.37$ and a
total mass $M = (11\,000 \pm 2600) \mbox{ M}_\odot$.  The error is
mainly due to the uncertainty on the distance of the cloud.  Note that
in this analysis we only included measurements above $b > +3^\circ$
(see discussion above in Sect. \ref{sec:nicer-absorpt-map}).

Our mass estimate apparently compares well with the independent one of
\citet{1999PASJ...51..871O}, $M \simeq 10^4 \mbox{ M}_\odot$.  Note,
however, that if we use the cloud distance assumed by
\citet{1999PASJ...51..871O}, $d = 160 \mbox{ pc}$, we obtain a larger
mass, $M = 1.7 \times 10^4 \mbox{ M}_\odot$, i.e., the CO derived mass
is only about 65\% the dust derived mass.  As discussed in
Sect.~\ref{sec:comparison-with-co}, this discrepancy can in principle
be attributed to (1) the insensitivity of $^{12}$CO to low column
densities, (2) to the saturation of $^{12}$CO in the dense cores of
the cloud, and (3) to a relatively small X-factor used by
\citet{1999PASJ...51..871O}. We can rule out the latter as the
X-factor used by \citet{1999PASJ...51..871O} ($2.8 \times 10^{20}
\mbox{ cm}^{-2} \mbox{ K}^{-1} \mbox{ km}^{-1} \mbox{ s}$) is
virtually coincident (96\%) with the one derived in this paper ($2.91
\times 10^{20} \mbox{ cm}^{-2} \mbox{ K}^{-1} \mbox{ km}^{-1} \mbox{
  s}$).

\begin{figure}[!tbp]
  \begin{center}
    \includegraphics[bb=128 298 456 530, width=\hsize]{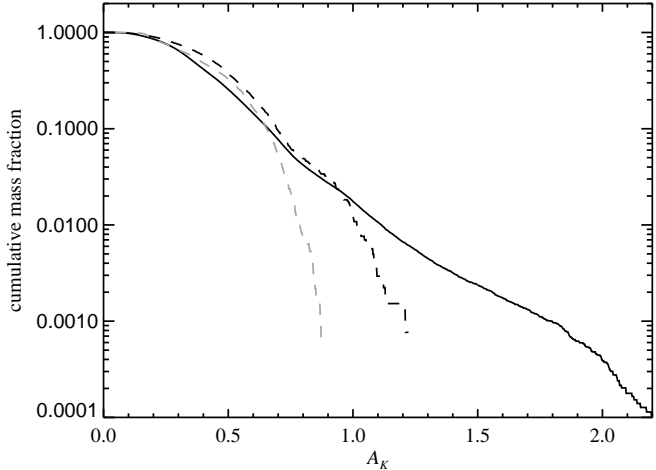}
    \caption{The cumulative mass enclosed in isoextinction contours.
      The plot has been constructed using only extinction measurements
      at galactic latitude $b > 3^\circ$ (cf.\ Fig.~\ref{fig:20}).
      The solid line shows the cumulative mass from the whole
      \textsc{Nicer} map, the dashed black line the cumulative mass
      from the downgraded (and clipped) map of Fig.~\ref{fig:21}
      right, and the dashed grey line the cumulative mass derived from
      the NANTEN observations (see Fig.~\ref{fig:21} right).  For this
      latter mass estimate we used the conversion formula
      \eqref{eq:15}.}
    \label{fig:26}
  \end{center}
\end{figure}

To investigate the source of discrepancy in the mass estimates we
present in Fig.~\ref{fig:26} the relationship between the integrated
mass distribution and the extinction in $A_K$. We also plot, as a
dashed line, the smoothed and clipped extinction map that was compared
to the CO data (see Fig.~\ref{fig:21}). Note that regions with
extinction larger than $A_K > 0.6$ magnitudes (where the CO-dust
correlation breaks at higher column densities) account for about
$20\%$ of the total mass (dashed line).  Similarly, note how regions
with $A_K < 0.25 \mbox{ mag}$, the column density threshold below
which CO is not sensitive to H$_2$, account for about $15\%$ of the
cloud mass. The total fraction of dust mass that is missed by the CO
is then about 35\%, which, within the approximations, solves the
discrepancy found between the dust and CO derived mass.

\section{Conclusions}
\label{sec:conclusions}

The main results of this paper can be summarized as follows:
\begin{itemize}
\item We used 4.5 million stars from the 2MASS point source catalog to
  construct a $8^\circ \times 6^\circ$ extinction map of the poorly
  studied Pipe nebula, a molecular cloud complex seen in projection
  towards the galactic bulge.  The map has a resolution of 1~arcmin and
  has a $3 \sigma$ detection level of $0.5$ visual magnitudes.
\item We combined the 2MASS data with Tycho and Hipparcos parallaxes
  to obtain a distance of $130 \pm 15 \mbox{ pc}$ to this molecular
  complex. We estimated, for this distance, a total mass of $\sim 10^4
  \mbox{ M}_\odot$ for the Pipe complex.  This makes the Pipe nebula
  one the closest star forming region to the Earth of its type, closer
  than the Ophiuchus and the Taurus complexes.
\item We studied in details the statistical properties of the
  extinction map obtained through the \textsc{Nicer} method.  We also
  confirmed the relation originally observed by
  \citet{1994ApJ...429..694L} for IC~5146, suggesting that there is
  unresolved structure at the highest extinctions in our map.
\item We compared our near-infrared study with the CO observations of
  \citet{1999PASJ...51..871O}, and derived fitting formulae to relate
  the $^{12}$CO column density with the $A_K$ extinction.  We also
  derived the dust-to-$^{12}$CO ratio, and showed that if a normal
  infrared reddening law is assumed, then the derived X-factor is as
  large as $2.91 \times 10^{20} \mbox{ cm}^{-2} \mbox{ K}^{-1} \mbox{
    km}^{-1} \mbox{ s}$ in the range $A_K \in [0.1, 0.6] \mbox{ mag}$.
\item We found that $^{12}$CO is only sensitive to about 65\% of the
  total dust mass. About half of the missing mass is not traced by CO
  at column densities below $A_K < 0.25 \mbox{ mag}$, and half is not
  traced at column densities above $A_K > 0.6 \mbox{ mag}$, where the
  line begins to saturate.  There is an apparently irreducible
  uncertainty in the dust-CO correlation of about 1 magnitude of
  visual extinction. This uncertainty seems to be independent of the
  cloud or the interstellar radiation field.
\item We took advantage of the large number of background sources to
  accurately measure the NIR reddening law for this cloud, and
  obtained $E(J - H) = (1.85 \pm 0.15) E(H - K)$, in very good
  agreement with the standard \citet{1985ApJ...288..618R} reddening
  law.
\item Finally from analysis of the JHK color-color diagram for the
  Pipe region we identify a large population of red stars whose colors
  are distinct from those of typical reddened background giants. These
  stars are spatially distributed over the entire observed field,
  preferentially located at lower Galactic latitudes and are not
  associated with the molecular cloud.  These stars are the brightest
  stars detected in the field and have a very narrow distribution in
  magnitude space.  These properties are similar to those of Galactic
  OH/IR stars.  Our observations thus appear to have provided one of
  the largest samples of such stars yet discovered.
\end{itemize}

\acknowledgements 

We thank Jerry Lodriguss for generously supplying large-field color
images of the Pipe nebula, and \citeauthor{1999PASJ...51..871O} for
kindly providing the $^{12}$CO NANTEN data.  This research has made
use of the 2MASS archive, provided by NASA/IPAC Infrared Science
Archive, which is operated by the Jet Propulsion Laboratory,
California Institute of Technology, under contract with the National
Aeronautics and Space Administration.  This paper also made use of the
Hipparcos and Tycho Catalogs (ESA SP-1200, 1997), the All-sky Compiled
Catalogue of 2.5 million stars (ASCC-2.5, 2001), and the Tycho-2
Spectral Type Catalog (2003).  CJL acknowledges support from NASA
ORIGINS Grant NAG 5-13041.

\bibliographystyle{aa} 
\bibliography{../dark-refs}

\end{document}